\documentclass[prd,superscriptaddress,altaffilletter,nofootinbib]{revtex4}

\usepackage{cancel}
\usepackage{graphicx}
\usepackage{amsmath}
\usepackage{amssymb}
\usepackage{graphicx,epsfig}
\usepackage[dvipsnames]{xcolor}
\newcommand{\be}{\begin{equation}}
\newcommand{\ee}{\end{equation}}
\newcommand{\bea}{\begin{eqnarray}}
\newcommand{\eea}{\end{eqnarray}}
\newcommand{\der}{\partial}
\newcommand{\vphi}{\varphi}

\begin{document}



\title{The Unknown Face of Scalar-Tensor Gravitational Theories}

\author{Israel Quiros}\email{iquiros@fisica.ugto.mx;i.quiros@ugto.mx}\affiliation{Departamento Ingenier\'ia Civil, Divisi\'on de Ingenier\'ia, Universidad de Guanajuato, C.P. 36000, Gto., M\'exico.}

\author{Amit Kumar Rao}\email{amit.kumar@ugto.mx;  amit.akrao@gmail.com}\affiliation{Departamento Ingenier\'ia Civil, Divisi\'on de Ingenier\'ia, Universidad de Guanajuato, C.P. 36000, Gto., M\'exico.}\affiliation{Instituto de F\'{\i}sica, Benem\'erita Universidad Aut\'onoma de Puebla,\\
Apartado Postal J-48, 72570, Puebla, Puebla, M\'exico.}

\date{\today}

\begin{abstract} It has long been demonstrated that the vacuum scalar-tensor theory in the Jordan-frame Brans-Dicke parametrization is form-invariant under conformal transformations, provided that a suitable transformation of the coupling parameter $\omega$ is applied. Here, we generalize this framework to include the coupling of matter fields to gravity. We take into consideration the recent result that, for point-dependent masses transforming as $m\rightarrow\Omega^{-1}m$ under the conformal transformations, the Lagrangian density of fundamental matter fields and perfect fluids is conformal form-invariant. We demonstrate that the conformal frame issue, that arises in the context of scalar-tensor gravity theories, is a consequence of two factors: i) the omission of the transformation of field-dependent parameters, such as the coupling function $\omega=\omega(\phi)$, under the conformal transformation of the fields, and ii) the ignorance of the Ward identity due to conformal form-invariance of the Lagrangian density of matter, which leads to an incorrect Klein-Gordon-type equation of motion for the Brans-Dicke field. By considering the conformal transformations as coordinate transformations in the configuration space, where the metric $g_{\mu\nu}$, the Brans-Dicke scalar $\phi$, and $N$ matter fields $\chi=\{\chi_1,\chi_2,...,\chi_N\}$, which are coupled to gravity, are assumed as ``generalized coordinates,'' we introduce the notion of active and passive conformal transformations. We demonstrate that passive conformal transformations do not represent a suitable framework for exploring the physical consequences of conformal symmetry; in contrast, active conformal transformations do.\end{abstract}



\maketitle




\section{Introduction}
\label{sect-intro}


The conformal frame issue (CFI) dates back to the early 1960s, after the well-known paper by Dicke \cite{dicke-1962}. The issue arises from different, sometimes opposite, interpretations of the conformal transformation (CT) of the metric \cite{dicke-1962, morganstern_1970, anderson-1971, bekenstein_1980, cotsakis_1993, magnano_1994, capozziello_1997, kaloper_prd_1998, faraoni_rev, faraoni_ijmpd_1999, quiros_prd_2000, fabris_2000, casadio_2002, alvarez_2002, flanagan_cqg_2004, bhadra_2007, hassaine_2007, fujii_2007, faraoni_prd_2007, catena_prd_2007, jarv_2007, sotiriou_ijmpd_2008, odi-1, elizalde_grg_2010, deruelle_2011, marques-2012, chiba_2013, capozziello_prd_2013, quiros_grg_2013, morris-2014, sasaki_2016, banerjee_2016, pandey_2017, azri-2018, rinaldi-2018, pal-2019, singleton-2020, quiros_ijmpd_2020, gionti_2021, pal_2022, bamber_prd_2023, moinuddin_epjc_2023, mukherjee_epjc_2023, mazumdar_prd_2023, bamba_prd_2024, gionti_epjc_2024, paliatha-2024, mandal_2024, dahal-2024}. Mathematically, CT can be stated as the following transformation of the metric tensor $g_{\mu\nu}$ and other fields $\Phi_i$ ($i=1, 2,..., N$) present in the theory:

\begin{align} g_{\mu\nu}\rightarrow\hat g_{\mu\nu}=\Omega^2g_{\mu\nu},\;\Phi_i\rightarrow\hat\Phi_i=\Omega^{w_i}\Phi_i,\label{conf-t}\end{align} where the positive smooth function $\Omega=\Omega(x)$ is known as the conformal factor and $w_i$ is the conformal weight of the field $\Phi_i$. This transformation acts only on fields, so it does not affect the spacetime coordinates and points; that is, this is not a spacetime diffeomorphism.


From a mathematical point of view \eqref{conf-t} is an equivalence relationship, so the different conformal representations of a given theory are mathematically equivalent. The question is: are these conformal representations physically equivalent? Seeking an answer to the latter question led to the CFI in the first place. The different points of view on the role that CT \eqref{conf-t} plays in scalar-tensor gravitational (STG) theories \cite{brans-dicke-1961, bergmann_1968, nordvedt_1970, wagoner_1970, anderson-1971, ross_1972, fujii-1974, isenberg_1976, casas_1992, fujii_book, faraoni_book, quiros_ijmpd_2019}, can be roughly summarized in the following way:\footnote{In this document, we consider ``traditional'' STG theories, leaving Horndeski theories for future work.}

\begin{itemize}

  \item The CT relates complementary, physically equivalent descriptions of gravitational laws \cite{dicke-1962, anderson-1971, marques-2012, morris-2014, pal-2019, paliatha-2024, fujii_book, faraoni_book, alvarez_2002, flanagan_cqg_2004, hassaine_2007, fujii_2007, faraoni_prd_2007, catena_prd_2007, sotiriou_ijmpd_2008, deruelle_2011, chiba_2013, sasaki_2016, pandey_2017, shtanov-2022, bamber_prd_2023, moinuddin_epjc_2023, mukherjee_epjc_2023}. These descriptions are usually called conformal frames. Among these, the Jordan frame (JF) and Einstein frame (EF) are distinguished.
	
  \item The different representations or conformal frames are not physically equivalent \cite{morganstern_1970, cotsakis_1993, magnano_1994, capozziello_1997, rinaldi-2018, paliatha-2024, kaloper_prd_1998, faraoni_rev, faraoni_ijmpd_1999, quiros_prd_2000, fabris_2000, casadio_2002, bhadra_2007, elizalde_grg_2010, capozziello_prd_2013, quiros_grg_2013, banerjee_2016, quiros_ijmpd_2020, gionti_2021, pal_2022, mazumdar_prd_2023, bamba_prd_2024, gionti_epjc_2024, mandal_2024}. It makes sense to wonder what conformal frame is the physical one. In particular, there are arguments in favor of the EF as well as in favor of the JF. In this case, CT can be used exclusively as a mathematical tool to simplify the mathematical handling of the equations of a given STG theory.
	
\end{itemize} 

For a detailed discussion of the different approaches to CT found in the bibliography and their role in STG theories, we recommend the review paper \cite{faraoni_rev}. In this well-known paper, the degree of overall confusion existing in the early 2000s about the CT and the frames issue is properly established. Unfortunately, almost nothing has changed on this issue since then. This is evident from the most recent bibliography on the subject \cite{flanagan_cqg_2004, bhadra_2007, hassaine_2007, fujii_2007, faraoni_prd_2007, catena_prd_2007, jarv_2007, sotiriou_ijmpd_2008, odi-1, elizalde_grg_2010, deruelle_2011, marques-2012, chiba_2013, capozziello_prd_2013, quiros_grg_2013, morris-2014, sasaki_2016, banerjee_2016, pandey_2017, azri-2018, rinaldi-2018, pal-2019, singleton-2020, quiros_ijmpd_2020, gionti_2021, pal_2022, bamber_prd_2023, moinuddin_epjc_2023, mukherjee_epjc_2023, mazumdar_prd_2023, bamba_prd_2024, gionti_epjc_2024, paliatha-2024, mandal_2024, dahal-2024}.


The origin of the CFI can be related to the very concept of "physical equivalence." There are not many publications in the bibliography that approach the question of how to define it \cite{qadir_1976, fatibene_2014}. Based on the few existing publications and on the usual meaning attributed to physical equivalence in several contexts, one may conclude that there is no consensus: Each researcher assigns a biased interpretation of physical equivalence, shaped by their approach to the conformal frame issue. 

A widespread understanding of physical equivalence is associated with form-invariance of the Lagrangian and of the derived equations of motion (EOM), with respect to certain local transformations or a group of transformations. If we follow this understanding and if the transformation of the field-dependent parameters is omitted, then the different conformal frames in which STG theories can be formulated are not physically equivalent, since these are characterized by Lagrangian densities that differ in the form in which they depend on the fields and, consequently, by different EOM. Just compare the JF gravitational Lagrangian density of STG theory in Jordan frame Brans-Dicke (JFBD) parametrization;

\begin{align} {\cal L}_\text{jfbd}=\frac{\sqrt{-g}}{2}\left[\phi R-\frac{\omega(\phi)}{\phi}(\der\phi)^2-2V(\phi)\right],\label{jfbd-lag}\end{align} where the BD scalar $\phi$ has dimensions of mass squared, with the EF one;

\begin{align} {\cal L}_\text{efbd}=\frac{M^2_\text{pl}}{2}\sqrt{-\hat g}\left[\hat R-\left(\frac{3}{2}+\omega\right)\frac{(\hat\der\phi)^2}{\phi^2}-2\hat V\right],\label{ef-lag}\end{align} which is obtained from \eqref{jfbd-lag} after the application of the following ``restricted conformal transformations'' (RCTs):

\begin{align} g_{\mu\nu}\rightarrow\hat g_{\mu\nu}=\Omega^2g_{\mu\nu},\;\phi\rightarrow\hat\phi=\Omega^{-2}\phi,\;V\rightarrow\hat V=\Omega^{-4}V,\label{gauge-t}\end{align} where the Einstein frame corresponds to the the choice $\hat\phi=M^2_\text{pl}$; that is, $\phi\rightarrow M^2_\text{pl}=\Omega^{-2}\phi$, so the self-interacting potential $\hat V=\Omega^{-4}V(\phi)/M^2_\text{pl}=M^2_\text{pl}V(\phi)/\phi^2$. 

Alternatively, as shown in \cite{faraoni_1998}, the STG Lagrangian density in the JFBD parametrization \eqref{jfbd-lag} is form-invariant under generalized conformal transformations (GCT) that include, in addition to CT \eqref{conf-t}, and transformations of the BD field $\phi\rightarrow\hat\phi=\Omega^{-2}\phi$, and of the self-interaction potential $V=V(\phi)$; $V\rightarrow\hat V=\Omega^{-4}V$, the following transformation of the coupling function $\omega=\omega(\phi)$; 

\begin{align} \omega\rightarrow\hat\omega=\frac{\omega+6\frac{\Omega_{,\phi}}{\Omega}\phi\left(1-\frac{\Omega_{,\phi}}{\Omega}\phi\right)}{\left(1-2\frac{\Omega_{,\phi}}{\Omega}\phi\right)^2}.\label{w-gauge-t}\end{align}


In the present document, for definiteness and philosophical prejudice, we associate physical equivalence with a symmetry group, so that the overall Lagrangian density and the derived EOM must be form-invariant under the group's transformations. As a symmetry group, we consider form-invariance under GCTs; 

\begin{align} g_{\mu\nu}\rightarrow\hat g_{\mu\nu}=\Omega^2g_{\mu\nu},\;\phi\rightarrow\hat\phi=\Omega^{-2}\phi,\;V\rightarrow\hat V=\Omega^{-4}V,\;\omega\rightarrow\hat\omega=\frac{\omega+6\frac{\Omega_{,\phi}}{\Omega}\phi\left(1-\frac{\Omega_{,\phi}}{\Omega}\phi\right)}{\left(1-2\frac{\Omega_{,\phi}}{\Omega}\phi\right)^2},\label{gen-gauge-t}\end{align} that include a conformal transformation of the metric \eqref{conf-t}, complemented by simultaneous transformations of other fields present, as well as parameters that depend on fields, such as the self-interaction potential and the coupling function (note that it is implicitly assumed that $\Omega=\Omega(\phi)$). In addition, we assume that the space-time points and their coordinates, the Planck constant $h$, the Planck mass $M^2_\text{pl}$, as well as any other constants, regardless of whether they are fundamental or not or whether dimensionless or with physical dimensions (this includes any integration constants), are not transformed by the GCTs. Only point-dependent fields are transformed by  \eqref{gen-gauge-t}. The EOMs derived from \eqref{jfbd-lag} are also unchanged by GCT, so STG theory has this symmetry group. This was demonstrated long ago in \cite{faraoni_1998} for the vacuum BD theory. Here, we generalize this result to the case when any matter fields are coupled to gravity. As we shall see, it is necessary to add to the GCT \eqref{gen-gauge-t} the following transformation of the mass parameter:

\begin{align} m\rightarrow\hat m=\Omega^{-1}m.\label{mass-t}\end{align} The transformation of the mass is required by dimensional analysis if CTs are to be interpreted as transformations of physical units in Dicke's sense \cite{dicke-1962}. In contrast, if we assume that the mass parameter is not transformed by the CTs, the resulting transformations are known as ``local scale transformations'' \cite{deser_1970} (see the clear discussion on the difference between transformations of units and local scale transformations in \cite{bekenstein_1980}). The transformation law of the masses \eqref{mass-t} is usually assumed when discussing conformal transformations in the framework of STG theories \cite{dicke-1962, bekenstein_1980, faraoni_rev, faraoni_prd_2007, quiros_ijmpd_2020, brans-dicke-1961, fujii_book, faraoni_book, quiros_ijmpd_2019, hobson-prd-2020, hobson-epjc-2022}. However, its actual role in the resolution of the CFI has been underestimated.\footnote{Note that the transformations of the mass $m$ and the Planck constant $h$ under conformal transformations are not independent. The Planck constant has the following dimensions: $[h]=[M][L]^2[T]^{-1}$, where $[M]$ is the mass dimension, while $[L]$ and $[T]$ are the length and time dimensions, respectively. Under conformal transformations, the latter dimensions transform as the proper length element $dl=\sqrt{g_{ik}dx^idx^k}$ and the proper time element $d\tau=\sqrt{g_{00}dt^2}$ do: $dl\rightarrow\Omega\,dl$ and $d\tau\rightarrow\Omega\,d\tau$. Consequently, $[L]^2[T]^{-1}\rightarrow\Omega\,[L]^2[T]^{-1}$. Hence, if we assume that CTs do not transform the Planck constant, the mass parameter $m$ must be transformed according to \eqref{mass-t}. Otherwise, if we assume that the mass parameter is a constant, then the Planck constant transforms as $h\rightarrow\Omega\,h$. Consequently, we have two possible transformation properties of the mass parameter $m$ under CTs: a) $m=$ const., which is not transformed: $m\rightarrow m$, and b) $m=m(x)$ is a point-dependent field that transforms according to \eqref{mass-t}. Here, we adopt the latter possibility, so that the Planck constant (as well as any other constant) is not transformed under CTs.} An important consequence of transformation \eqref{mass-t} is that, under CT \eqref{conf-t}, the energy density $\rho$ (also pressure) of perfect fluids, which has the following units: $[\rho]=[M]/[L]^3$, transforms as $\rho\rightarrow\hat\rho=\Omega^{-4}\rho$. Yet, as we shall show, its most important consequence is that the transformation \eqref{mass-t} is the necessary and sufficient condition for the Lagrangian density of matter to be form-invariant under GCTs \eqref{gen-gauge-t}.


When matter fields are coupled to gravity in STG-JFBD theory \eqref{jfbd-lag}, there is a subtlety associated with the variational derivative of the Lagrangian density of matter fields ${\cal L}_m={\cal L}_m(\chi,\der\chi,g_{\mu\nu})=\sqrt{-g}\,L_m(\chi,\der\chi,g_{\mu\nu})$, where $L_m$ is the matter Lagrangian and $\chi$ is the collective name for a set of $N$ matter fields $\chi=\{\chi_1,\chi_2,...,\chi_N\}$, which are coupled to gravity: If the masses of timelike fields are point-dependent fields which, under conformal transformations, transform as in \eqref{mass-t}, then the dimensional analysis yields the following result: the energy density of perfect fluids transforms as $\rho\rightarrow\hat\rho=\Omega^{-4}\rho$ (the same is true for the pressure of the fluid). Besides, since under conformal transformations the mass parameter transforms as the square root of the scalar field $\sqrt\phi$ does; that is, $m\propto\sqrt\phi$, one can assume for the masses $m_i$ of any timelike fields that, $m_i(x)=\kappa_i\sqrt{\phi(x)}$, where $\kappa_i$ are dimensionless constants \cite{hobson-prd-2020, hobson-epjc-2022}. Hence, for timelike fields, the Lagrangian density of matter depends on the BD field $\phi$ through the masses $m=m(\phi)$, while for perfect fluids it depends on $\phi$ through the energy density $\rho=\rho(\phi)$. That is, whenever the masses transform like $m\to\Omega^{-1}m$, the variational derivative of the Lagrangian density of matter with respect to the scalar field is nonvanishing; $\delta{\cal L}_m/\delta\phi\neq 0$. On the other hand, due to conformal form-invariance of the Lagrangian density of matter, under infinitesimal conformal transformations:

\begin{align} \delta g_{\mu\nu}=2\theta\,g_{\mu\nu}\,\left(\delta g^{\mu\nu}=-2\theta\,g^{\mu\nu},\;\delta\sqrt{-g}=4\theta\sqrt{-g}\right),\;\delta\phi=-2\theta\,\phi,\label{inf-gauge-t}\end{align} where for convenience (and only temporarily) we have rescaled the conformal factor: $\Omega=e^\theta$, the overall variation of the Lagrangian density of matter satisfies $\delta{\cal L}_m=0$ (or, in general, $\delta{\cal L}_m=\nabla_\mu V^\mu$, where $V^\mu$ is some vector). We have, $\delta S_m=\int d^4x\,\delta{\cal L}_m=0$, so that under \eqref{inf-gauge-t}, $S_m\rightarrow S_m+\delta S_m=S_m$; that is, the matter action is invariant under \eqref{inf-gauge-t}. For the overall variation of the Lagrangian density of matter, we have\footnote{In classical gravitational problems, it is understood that the matter fields are apriori given; that is, the EOMs of the matter fields: $$\frac{\delta{\cal L}_m}{\delta\chi_i}=0\;\Rightarrow\;\frac{\der{\cal L}_m}{\der\chi_i}-\nabla_\mu\left[\frac{\der{\cal L}_m}{\der(\der_\mu\chi_i)}\right]=0,$$ are automatically satisfied.}

\begin{align} \delta{\cal L}_m=\frac{\delta{\cal L}_m}{\delta g^{\mu\nu}}\delta g^{\mu\nu}+\frac{\delta{\cal L}_m}{\delta\phi}\delta\phi=0\;\Rightarrow\;-2\theta\left(g^{\mu\nu}\frac{\delta{\cal L}_m}{\delta g^{\mu\nu}}+\phi\frac{\delta{\cal L}_m}{\delta\phi}\right)=0.\label{var-lmat}\end{align} Above, to get from the equation on the left to the one on the right, we substituted $\delta g^{\mu\nu}=-2\theta\,g^{\mu\nu}$ and $\delta\phi=-2\theta\,\phi$ from the infinitesimal conformal transformations \eqref{inf-gauge-t}. From the equation on the right in \eqref{var-lmat}, it follows the Ward identity \cite{alvarez-2015}:

\begin{align} g^{\mu\nu}\frac{\delta{\cal L}_m}{\delta g^{\mu\nu}}=-\phi\frac{\delta{\cal L}_m}{\delta\phi}.\label{mat-ward-id}\end{align} This equation, which plays an important role in the derivation of the correct equations of motion when conformal invariance is a symmetry of the theory, has been overlooked in the literature on STG theories and the CFI.


This paper presents a new perspective on the frames issue. Our analysis will be based on two facts: i) form-invariance of the Lagrangian density under GTC \eqref{gen-gauge-t} plus simultaneous transformation of the mass \eqref{mass-t}, and ii) proper consideration of the Ward identity \eqref{mat-ward-id} associated with conformal form-invariance of the Lagrangian density of matter. In addition, we make a clear distinction between the two complementary approaches to conformal transformations \cite{quiros-prd-2025}: 1) passive approach to conformal transformations (PACT) and 2) active approach to conformal transformations (AACT). These are a natural result of considering CT as a coordinate transformation on the configuration space manifold ${\cal M}_\text{fields}$, where the different fields, including the scalar field, the metric, and the matter fields, are treated as ``generalized coordinates''. The idea of considering the scalar fields as generalized coordinates in the configuration space has been previously proposed in \cite{vilko-npb-1984} to search for a unique effective action in quantum field theory (QFT) and also in \cite{gong_2011} in the context of multifield inflation, by assuming that the multiple scalar fields are coordinates in some multidimensional field-space manifold (see also \cite{stein-2013, jarv_2015, karamitsos_2016, karamitsos_2018, karamitsos_2020, karamitsos_2021}). As stated in \cite{gong_2011}, it is preferable to formulate the dynamics in the configuration space in a coordinate-independent (covariant) manner. In this document, we shall further generalize the aforementioned formalism to consider any fields: scalar fields, metric, and matter fields, as generalized coordinates in the configuration space. This generalization allows us to find that there are two possible approaches to conformal transformations: PACT and AACT. The latter is required to explain the possible physical and phenomenological impact of conformal symmetry. 

We pay special attention to the coupling of matter fields with gravity. The result recently demonstrated in \cite{quiros-prd-2025}, that if we regard the masses as point-dependent fields which under \eqref{conf-t} transform as in \eqref{mass-t}, the Lagrangian density of fundamental matter fields and perfect fluids is conformal form-invariant, leads to the following question:\footnote{The demonstration of conformal form-invariance of the action for fermion fields has been given in \cite{bekenstein_1980} and can also be found in \cite{sasaki_2016}. However, in none of these bibliographic references was the importance and generality of the demonstration discussed. The novel aspect of the demonstration of conformal form-invariance of the Lagrangian of matter fields in \cite{quiros-prd-2025} is that the Lagrangian density of QCD fermions, as well as gauge fields in general, is included. In addition, the demonstration in \cite{quiros-prd-2025} and in this document includes the Lagrangian density of perfect fluids. This marks a huge difference with \cite{bekenstein_1980, sasaki_2016} since the generality of the conformal form-invariance of the Lagrangian density of matter fields is highlighted.} Why couldn't the gravitational fields be conformal form-invariant as well? This question is at the core of the CFI: The total Lagrangian density for STG theory in JFBD parametrization is given by the following expression: ${\cal L}_\text{tot}={\cal L}_\text{jfbd}+{\cal L}_m$, where ${\cal L}_\text{jfbd}$ is given by \eqref{jfbd-lag}. The problem is that, under CTs: ${\cal L}_m\to{\cal L}_m$, while ${\cal L}_\text{jfbd}\cancel{\to}{\cal L}_\text{jfbd}$. That is, the gravitational Lagrangian density spoils conformal form-invariance of the overall Lagrangian density.

In this document, we will demonstrate that the existing discrepancy between the fact that the gravitational Lagrangian density of STG theory in JFBD parametrization is not conformal form-invariant, while the Lagrangian density of matter fields is indeed conformal form-invariant, is due to the omission of the transformation \eqref{w-gauge-t} of the coupling parameter $\omega=\omega(\phi)$ under conformal transformations. The coupling parameter depends on the BD scalar field $\phi$, which is transformed under conformal transformations: $\phi\rightarrow\hat\phi=\Omega^{-2}\phi$. Hence, $\omega$ is expected to transform accordingly. The omission of the transformation of the coupling parameter is at the heart of the CFI. In fact, it has been known for a long time \cite{faraoni_1998}, that if we add a suitable transformation of the coupling parameter \eqref{w-gauge-t} to the conformal transformation \eqref{conf-t}, the vacuum STG in the JFBD parametrization is not only conformal invariant but also form-invariant. Here, we generalize this result to the case where matter fields are coupled to gravity. 


In the present document, we address, basically, Dicke's hypothesis that ``the laws of physics must be invariant under a transformation of units'' \cite{dicke-1962}, which we identify with the following GCTs:

\begin{align} g_{\mu\nu}&\rightarrow\hat g_{\mu\nu}=\Omega^2g_{\mu\nu},\;\phi\rightarrow\hat\phi=\Omega^{-2}\phi,\;V\rightarrow\hat V=\Omega^{-4}V,\;\chi\rightarrow\hat\chi=\Omega^{w_\chi}\chi,\nonumber\\
\omega&\rightarrow\hat\omega=\frac{\omega+6\frac{\Omega_{,\phi}}{\Omega}\phi\left(1-\frac{\Omega_{,\phi}}{\Omega}\phi\right)}{\left(1-2\frac{\Omega_{,\phi}}{\Omega}\phi\right)^2},\;m\rightarrow\hat m=\Omega^{-1}m,\label{gct}\end{align} where $w_\chi$ is the conformal weight of the matter field $\chi$.

We must agree that the physical laws, in particular the gravitational laws, are expressed through the EOM. Hence, it is reasonable to assume that Dicke's hypothesis entails that the EOM of STG theory must be form-invariant under GCTs \eqref{gct}. This is possible only if the total Lagrangian density is not only invariant but also form-invariant under these generalized conformal transformations. In what follows, we refer to Dicke's hypothesis, complemented by the identification of the physical laws and the EOM, as "Dicke's principle." We shall show that Dicke's principle is satisfied by STG theories only if we consider the generalized CT \eqref{gct}, and if, simultaneously, we take into account the Ward identity \eqref{mat-ward-id}, which is associated with conformal form-invariance of the Lagrangian density of matter fields. Hence, fulfillment of Dicke's principle implies removal of CFI.


The paper is organized as follows. The basic mathematical and physical aspects of scalar-tensor gravitational theory in the JFBD parametrization, as well as the nomenclature and symbology, are reviewed in Section \ref{sect-stg}. The conformal form-invariant formulation of STG theory is investigated in Section \ref{sect-jfbd-par}. In this section, the Ward identity associated with conformal form-invariance of the Lagrangian density of matter \eqref{mat-ward-id} is taken into account during the application of the variational procedure, and the correct EOM are derived. In Section \ref{sect-fiss}, we explain the origin of the CFI in STG theories as a result of the incorrect omission of the transformation \eqref{w-gauge-t} of the coupling parameter $\omega$ under CT, and also due to ignorance of the Ward identity \eqref{mat-ward-id} associated with conformal form-invariance of the Lagrangian of matter fields; that is, ignorance of the transformation law \eqref{mass-t} of masses. In Section \ref{sect-act-pass}, we discuss active and passive approaches to CT. The discussion of this issue is necessary because, as we shall show, the passive conformal transformations (the most frequent approach in the bibliography) are not suitable for addressing the CFI. Only if we follow the active approach to conformal transformations can these have physical (also phenomenological) consequences. In the present document, we follow, precisely, the active approach to the conformal transformations. In Section \ref{sect-conf-space}, the STG theories are formulated in covariant form in the configuration space. We will show that only if we consider form-invariance under GCTs \eqref{gct}, the STG theory in JF parametrization can be written in a covariant way in the configuration space. In Section \ref{sect-gauges}, the notion of gauge freedom within the present framework, in which STG theories in the JFBD parametrization are treated as conformal form-invariant, and the active approach to CTs is adopted, is discussed. The representation of the resulting picture, which is similar to the many-worlds interpretation of quantum physics, is also discussed. In Section \ref{sect-extra}, we summarize the outstanding results and discuss the main phenomenological consequences of our current framework. We also compare the present results with the well-known results in the literature on scalar-tensor theories. Several cornerstone aspects of the conformal transformations issue are briefly reviewed and discussed in Section \ref{sect-discu}, where the main conclusions are also summarized.



\section{Scalar-Tensor Gravitational Theory}
\label{sect-stg}


The STG theories are given generically by the following Lagrangian density \cite{brans-dicke-1961, bergmann_1968, nordvedt_1970, wagoner_1970, ross_1972, fujii-1974, isenberg_1976, casas_1992, fujii_book, faraoni_book, quiros_ijmpd_2019}:

\begin{align} {\cal L}_\text{stg}=\frac{\sqrt{-g}}{2}\left[fR-w(\der\vphi)^2-2V+2L_m\right],\label{jf-action}\end{align} where $R$ is the curvature scalar, while the gravitational coupling $f=f(\vphi)$, the coupling parameter $w=w(\vphi)$ and the self-interaction potential $V=V(\vphi),$ are functions of the scalar field $\vphi$. In addition, $L_m=L_m(\chi,\der\chi,g_{\mu\nu})$ is the Lagrangian of the matter fields, which are collectively denoted by $\chi$, and we have used the following notation: $(\der\vphi)^2\equiv g^{\mu\nu}\der_\mu\vphi\der_\nu\vphi$. The Lagrangian density \eqref{jf-action} is said to be written in JF variables. However, this is not the only way to write the JF representation of STG theories. If one makes the following redefinition: $\phi=f(\vphi),$ where the scalar field $\phi$ has dimensions of mass squared, and defining $\omega(\phi)\equiv\phi(f^{-1}_{,\phi})^2w(\phi)$, the gravitational part of the JF Lagrangian density \eqref{jf-action} can be written alternatively in the JFBD parameterization \eqref{jfbd-lag}; ${\cal L}_\text{jfbd}=\sqrt{-g}\left[\phi R-\omega(\phi)(\der\phi)^2/\phi-2V(\phi)\right]/2$. BD theory \cite{brans-dicke-1961} is a particular case of \eqref{jfbd-lag} when the coupling function $\omega(\phi)=\omega_\text{BD},$ is a constant, and the self-interaction potential vanishes $V=0$. In what follows, when we refer to the JF Lagrangian density of a generic STG theory, we assume it is composed of the gravitational and matter Lagrangian densities.


From the overall Lagrangian density ${\cal L}_\text{tot}={\cal L}_\text{jfbd}+{\cal L}_m$, where ${\cal L}_\text{jfbd}$ is the JFBD Lagrangian density \eqref{jfbd-lag} and ${\cal L}_m=\sqrt{-g}\,L_m(\chi,\der\chi,g_{\mu\nu})$ is the Lagrangian density of matter fields, the following EOM may be derived: i) the Einstein-Brans-Dicke (EBD) EOM

\begin{align} G_{\mu\nu}=\frac{\omega}{\phi^2}\left[\der_\mu\phi\der_\nu\phi-\frac{1}{2}g_{\mu\nu}(\der\phi)^2\right]-g_{\mu\nu}\frac{V}{\phi}+\frac{1}{\phi}\left(\nabla_\mu\nabla_\nu-g_{\mu\nu}\nabla^2\right)\phi+\frac{1}{\phi}T^{(m)}_{\mu\nu},\label{ebd-eom}\end{align} where $\omega=\omega(\phi)$, $V=V(\phi)$, $G_{\mu\nu}=R_{\mu\nu}-g_{\mu\nu}R/2$ is the Einstein tensor, and $\nabla^2\equiv g^{\mu\nu}\nabla_\mu\nabla_\nu$ and ii) the Klein-Gordon-Brans-Dicke (KGBD) EOM

\begin{align} 2\omega\nabla^2\phi+\left(\omega_{,\phi}-\frac{\omega}{\phi}\right)(\der\phi)^2+\phi R=2\phi V_{,\phi},\label{kgbd-eom}\end{align} respectively. By inserting the trace of \eqref{ebd-eom}, the latter EOM can be written in the following equivalent form:

\begin{align} \left(3+2\omega\right)\nabla^2\phi+\omega_{,\phi}(\der\phi)^2=2\left(\phi V_{,\phi}-2V\right)+T^{(m)}.\label{kgbd-eom'}\end{align} In the above equations, we have used the following notation $X_{,\phi}\equiv\der X/\der\phi$, while $T^{(m)}=g^{\mu\nu}T^{(m)}_{\mu\nu}$ is the trace of the stress-energy tensor (SET) of matter, which is defined in the usual way:

\begin{align} T^{(m)}_{\mu\nu}=-\frac{2}{\sqrt{-g}}\frac{\delta{\cal L}_m}{\delta g^{\mu\nu}}.\label{m-set}\end{align} 

If we take the covariant divergence of \eqref{ebd-eom} and consider the second Bianchi identity ($\nabla^\mu G_{\mu\nu}=0$) and the identity $(\nabla_\mu\nabla_\nu-\nabla_\nu\nabla_\mu)\nabla^\mu\phi=R_{\nu\mu}\nabla^\mu\phi$, we get the following equation; 

\begin{align} \nabla^\lambda T^{(m)}_{\lambda\mu}=-\frac{\der_\mu\phi}{2\phi}\left[2\omega\nabla^2\phi+\left(\omega_{,\phi}-\frac{\omega}{\phi}\right)(\der\phi)^2+\phi R-2\phi V_{,\phi}\right].\label{ncons-eq}\end{align} Then, if we further consider the KGBD-EOM \eqref{kgbd-eom}, we obtain the JF conservation equation:

\begin{align} \nabla^\lambda T^{(m)}_{\lambda\mu}=0.\label{jf-cons-eq}\end{align} We see that the conservation equation \eqref{jf-cons-eq} is a consequence of the JFBD-EOM \eqref{ebd-eom} and \eqref{kgbd-eom}.



\section{Conformal form-invariance of scalar-tensor theory in Jordan frame Brans-Dicke parametrization with matter sources}
\label{sect-jfbd-par}


Let us now consider the conformal form-invariant STG theory in the JFBD parametrization, whose overall Lagrangian density reads:

\begin{align} {\cal L}_\text{tot}=\frac{\sqrt{-g}}{2}\left[\phi R-\frac{\omega}{\phi}(\der\phi)^2-2V\right]+{\cal L}_m(\chi,\der\chi,g_{\mu\nu}).\label{tot-lag}\end{align} As shown in \cite{faraoni_1998}, the gravitational Lagrangian density \eqref{jfbd-lag}, and the derived vacuum EOM, are form-invariant under GCT \eqref{gen-gauge-t}, which, in addition to CT \eqref{gauge-t}, includes a rescaling of the coupling parameter \eqref{w-gauge-t}. Actually, it is not difficult to prove that under GCT \eqref{gen-gauge-t} the gravitational Lagrangian density \eqref{jfbd-lag} is transformed into the following:

\begin{align} {\cal L}_\text{jfbd}=\frac{\sqrt{-\hat g}}{2}\left[\hat\phi\hat R-\frac{\hat\omega}{\hat\phi}(\hat\der\hat\phi)^2-2\hat V\right].\nonumber\end{align} That is, it is conformal form-invariant.


In \cite{quiros-prd-2025} it was shown that, if we assume the transformation \eqref{mass-t} of masses, the Lagrangian density of matter fields ${\cal L}_m={\cal L}_m(\chi,\der\chi,g_{\mu\nu})$ is conformal form-invariant; that is, under generalized conformal transformations \eqref{gct} plus the simultaneous transformation of matter fields $\chi\rightarrow\hat\chi=\Omega^{w_\chi}\chi$, the following transformation takes place: 

\begin{align} {\cal L}_m(\chi,\der\chi,g_{\mu\nu})\rightarrow{\cal L}_m(\hat\chi,\hat\der\hat\chi,\hat g_{\mu\nu})={\cal L}_m(\chi,\der\chi,g_{\mu\nu}).\label{lag-t}\end{align} Thanks to this transformation property, we can extend the form-invariance of the vacuum STG theory in the JFBD parametrization to the total Lagrangian density \eqref{tot-lag}.

Although it is not difficult to show that the vacuum gravitational Lagrangian \eqref{jfbd-lag} is form-invariant under GCT \eqref{gen-gauge-t} \cite{faraoni_1998}, when matter is coupled to gravity, as in \eqref{tot-lag}, the loss of conformal form-invariance seems to be inevitable. Actually, the EOM derived from \eqref{tot-lag} by the variational principle of least action are the Einstein-type equation \eqref{ebd-eom} and the KG-type equation \eqref{kgbd-eom} or its equivalent \eqref{kgbd-eom'}. The Einstein-type EOM is obviously form-invariant under GCT \eqref{gen-gauge-t} and should therefore be the KG-type equations \eqref{kgbd-eom} and \eqref{kgbd-eom'}, since all these equations are derived from the Lagrangian density \eqref{tot-lag}, which is itself form-invariant under GCT \eqref{gen-gauge-t}. However, when in the non-homogeneous continuity equation derived from these EOM \eqref{ncons-eq} we substitute the KGBD-EOM \eqref{kgbd-eom}, we obtain the standard conservation equation \eqref{jf-cons-eq}; $\nabla^\lambda T^{(m)}_{\lambda\mu}=0$, which is not conformal form-invariant, unless the trace of the matter SET vanishes $T^{(m)}=0$. That is, according to the standard procedure, only radiation can be consistently coupled to conformal form-invariant gravity. This result arises due to ignorance of the Ward identity \eqref{mat-ward-id}, which is associated with conformal form-invariance of the matter Lagrangian density ${\cal L}_m(\chi,\der\chi,g_{\mu\nu})$. In this section, we will correct this.



\subsection{Form-invariance of the matter Lagrangian density}
\label{subsect-matt-coup}


Form-invariance of the Lagrangian density of matter fields is not a trivial fact. It must be required that the masses of timelike fields be themselves point-dependent fields $m=m(x)$ which, under the GCT \eqref{gen-gauge-t}, transform as in \eqref{mass-t}; $m\rightarrow\hat m=\Omega^{-1}m,$ which is, precisely, the case considered within the framework of STG theories \cite{dicke-1962, bekenstein_1980, faraoni_rev, faraoni_prd_2007, quiros_ijmpd_2020, brans-dicke-1961, fujii_book, faraoni_book, quiros_ijmpd_2019, hobson-prd-2020, hobson-epjc-2022}. For example, one can assume that mass is a universal function of the scalar field: $m_i(x)=\kappa_i\sqrt{\phi(x)}$ \cite{hobson-prd-2020, hobson-epjc-2022}, where $\kappa_i$ are dimensionless constants. This choice is dictated by the fact that, since the scalar field $\phi$ has squared mass dimensions, under conformal transformations the mass parameter transforms as $\sqrt\phi$ does; that is, $m\propto\sqrt\phi$.


Let us demonstrate the transformation law \eqref{lag-t} in two important cases: a) a fundamental field (for illustration, here we consider the Dirac fermion, but gauge fields have also been considered in \cite{quiros-prd-2025}), and b) the perfect fluid. These cases exhaust the most illustrative examples of sources of gravity in classical gravity contexts.


The Lagrangian density of a Dirac fermion $\psi$ ($\bar\psi$ is the adjoint spinor of the fermion) in a curved space is given by:

\begin{align} {\cal L}_\text{dirac}=\sqrt{-g}\bar\psi\left(i\cancel{\cal D}+m\right)\psi,\label{dirac-lag}\end{align} where $\cancel{\cal D}:=\gamma^\mu{\cal D}_\mu$ and the following covariant derivative has been defined:

\begin{align} {\cal D}_\mu\psi=\left(D_\mu-\frac{1}{2}\sigma_{ab}e^{b\nu}\nabla_\mu e^a_\nu\right)\psi.\label{cov-der-psi}\end{align} In these equations $a,b,c=0,1,2,3$ are the flat spacetime indices, while $\gamma^a$ are the Dirac gamma matrices, $e^a_\mu$ are the tetrad fields such that $g_{\mu\nu}=\eta_{ab}e^a_\mu e^b_\nu$ ($\eta_{ab}$ is the Minkowski metric). Besides $\gamma^\mu=e^\mu_c\gamma^c$, and 

\begin{align} \sigma_{ab}=\frac{1}{2}\left[\gamma_a,\gamma_b\right]=\frac{1}{4}\left(\gamma_a\gamma_b-\gamma_b\gamma_a\right),\nonumber\end{align} are the generators of the Lorentz group in the spin representation. Above, we have used the standard definition of the $SU(2)\times U(1)$ gauge derivative, $D_\mu\psi$ (see, for example, Equation (32) of \cite{quiros-prd-2025}). It can be shown that, under GCT \eqref{gen-gauge-t} plus simultaneous transformation of the fermion spinor 

\begin{align} \psi\rightarrow\Omega^{-3/2}\psi,\;\bar\psi\rightarrow\Omega^{-3/2}\bar\psi,\label{fermion-t}\end{align} the following transformation takes place \cite{quiros-prd-2025, quiros-arxiv2-2025}: $\cancel{\cal D}\psi\rightarrow\Omega^{-5/2}\cancel{\cal D}\psi$, so that $\bar\psi\cancel{\cal D}\psi\rightarrow\Omega^{-4}\bar\psi\cancel{\cal D}\psi$. Then, since $\sqrt{-g}\rightarrow\Omega^4\sqrt{-g}$, the massless part of the fermion Lagrangian density is form-invariant:

\begin{align} i\sqrt{-g}\,\bar\psi\cancel{\cal D}\psi\rightarrow i\sqrt{-g}\,\bar\psi\cancel{\cal D}\psi.\nonumber\end{align} From \eqref{dirac-lag} it follows that, since $\bar\psi\psi\rightarrow\Omega^{-3}\bar\psi\psi$, form-invariance of the mass piece of the Dirac Lagrangian density: $\sqrt{-g}\,m\bar\psi\psi$, is achieved only if the mass parameter transforms as \eqref{mass-t}.


Let us show that the Lagrangian density of perfect fluids is form-invariant under GCT \eqref{gen-gauge-t}. For a perfect fluid with energy density $\rho$ and barotropic pressure $p$ the Lagrangian density can be written as \cite{hawking-book, berto-prd-2008, schutz_1970}

\begin{align} {\cal L}_\text{fluid}=-\sqrt{-g}\,\rho.\label{pf-lag}\end{align} Although it can also be written as ${\cal L}_\text{fluid}=\sqrt{-g}\,p$, unless the perfect fluid is explicitly coupled to the curvature, the latter Lagrangian density and \eqref{pf-lag} are equivalent \cite{faraoni_2009}. Here, for definiteness, we choose \eqref{pf-lag}. 

According to dimensional arguments, if we assume point-dependent masses that transform as $m\rightarrow\Omega^{-1}m$ under CT \eqref{conf-t}, the energy density of the fluid transforms as $\rho\rightarrow\Omega^{-4}\rho$. This is easily understood if one notices that the energy density has units: $[M]/[L]^3$, where $[M]\rightarrow\Omega^{-1}[M]$ and $[L]\rightarrow\Omega[L]$ are the transformation laws of the units of mass and length under CT, respectively. Hence, $[M]/[L]^3\rightarrow\Omega^{-4}[M]/[L]^3$. The barotropic pressure of the fluid $p$ transforms in the same way: $p\rightarrow\Omega^{-4}p$. Since under conformal transformation, $\sqrt{-g}\rightarrow\sqrt{-\hat g}=\Omega^4\sqrt{-g}$ and $\rho\rightarrow\hat\rho=\Omega^{-4}\rho$, then $-\sqrt{-g}\,\rho=-\sqrt{-\hat g}\,\hat\rho$. That is, the Lagrangian density of the perfect fluid is form-invariant under conformal transformations. In contrast, if we assume that the mass unit is not transformed, then, under CT: $\rho\rightarrow\hat\rho=\Omega^{-3}\rho$, which means that the Lagrangian density of the perfect fluid is not form-invariant: $-\sqrt{-g}\,\rho=-\sqrt{-\hat g}\,\Omega\hat\rho$. Hence, the conformal form-invariance of the Lagrangian density of matter fields, including that of perfect fluids, is correlated with the transformation $m\rightarrow\Omega^{-1}m$ of the masses \eqref{mass-t}.


In the bibliography on STG theories and conformal transformations, one frequently finds statements like this (see, for example, discussion on pages 2 and 3, and equations (2.6), (2.7) of Ref. \cite{sasaki-2015}): ``...why should we consider Jordan frame if the Einstein frame is much simpler? ... once we take matter into account, one can usually define a frame where matter is minimally coupled to the metric and, therefore, ..., physical interpretation in that frame is straightforward.'' According to the mentioned bibliographic references, the matter action can be written as\footnote{Here, we use a notation that differs a bit from that in \cite{sasaki-2015}.}

\begin{align} S_m=\int d^4x\sqrt{-g}\,L_m[\chi,g_{\mu\nu}]=\int d^4x\sqrt{-\hat g}\,\Omega^{-4}L_m[\chi,\Omega^{-2}\hat g_{\mu\nu}],\nonumber\end{align} where the expression on the left is in JF, while the one on the right is in EF variables. First, notice that the above equality means that the action is conformal invariant but not form-invariant. Second, it is apparent that the matter field $\chi$ is not transformed by CT \eqref{conf-t}, which is correct only for massless fields (radiation). In general, if one also includes timelike fields, the correct expression reads

\begin{align} S_m=\int d^4x\sqrt{-g}\,L_m[\chi,g_{\mu\nu}]=\int d^4x\sqrt{-\hat g}\,\Omega^{-4}L_m[\Omega^{-w_\chi}\hat\chi,\Omega^{-2}\hat g_{\mu\nu}],\label{sasaki}\end{align} where $w_\chi$ is the conformal weight of the matter field $\chi$. The missing argument in this analysis is that, if the mass parameters are assumed to transform as $m\to\hat m=\Omega^{-1}m$ under CT \eqref{conf-t}, as is commonly assumed in the bibliography on conformal transformations, then, as shown in \cite{quiros-prd-2025};

\begin{align} \int d^4x\sqrt{-\hat g}\,\Omega^{-4}L_m[\Omega^{-w_\chi}\hat\chi,\Omega^{-2}\hat g_{\mu\nu}]=\int d^4x\sqrt{-\hat g}\,L_m[\hat\chi,\hat g_{\mu\nu}],\nonumber\end{align} so that

\begin{align} S_m=\int d^4x\sqrt{-g}\,L_m[\chi,g_{\mu\nu}]=\int d^4x\sqrt{-\hat g}\,L_m[\hat\chi,\hat g_{\mu\nu}],\nonumber\end{align} which represents the same equivalence as in \eqref{lag-t}. That is, the matter action is not only conformal invariant but also form-invariant. Form-invariance is required for the equations of motion to be conformal form-invariant as well. 

This result is contrary to the argument cited above. While the matter field $\chi$ is minimally coupled to the JF metric $g_{\mu\nu}$, the conformal field $\hat\chi$ is minimally coupled to the EF metric $\hat g_{\mu\nu}$. Therefore, both metric fields are physical, i.e., the argument establishing that in the EF the matter is coupled to the conformal metric $\Omega^{-2}\hat g_{\mu\nu}$ so that the JF metric is the physical one is incorrect.



\subsection{Conformal form-invariant equations of motion}
\label{subsect-vprinciple}


The assumption that the masses are fields with conformal weight $w_m=-1$ leads to the Lagrangian density of matter fields being form-invariant under generalized conformal transformations \eqref{gct}. This means, in turn, that the Ward identity \eqref{mat-ward-id} takes place;

\begin{align} \frac{\delta{\cal L}_m}{\delta\phi}=\frac{\sqrt{-g}}{2\phi}\,T^{(m)},\label{usef-rel}\end{align} where we have substituted the definition \eqref{m-set} of the matter SET. 

Next, consider the variations of the total action: 

\begin{align} S_\text{tot}=\int d^4x ({\cal L}_\text{jfbd}+{\cal L}_m)=\frac{1}{2}\int d^4x\sqrt{-g}\left[\phi R-\frac{\omega}{\phi}(\der\phi)^2-2V\right]+\int d^4x\,{\cal L}_m(\chi,\der\chi,g_{\mu\nu}).\label{tot-action}\end{align} That is,

\begin{align} \delta S_\text{tot}=\delta S_\text{jfbd}+\delta S_m=\int d^4x&\left[\frac{\delta{\cal L}_\text{jfbd}}{\delta g^{\mu\nu}}\delta g^{\mu\nu}+\frac{\delta{\cal L}_\text{jfbd}}{\delta\phi}\delta\phi+\frac{\delta{\cal L}_m}{\delta g^{\mu\nu}}\delta g^{\mu\nu}+\frac{\delta{\cal L}_m}{\delta\phi}\delta\phi\right]=0,\nonumber\end{align} or, after rearrangement of terms:

\begin{align} \delta S_\text{tot}=&\int d^4x\left[\left(\frac{\delta{\cal L}_\text{jfbd}}{\delta g^{\mu\nu}}+\frac{\delta{\cal L}_m}{\delta g^{\mu\nu}}\right)\delta g^{\mu\nu}+\left(\frac{\delta{\cal L}_\text{jfbd}}{\delta\phi}+\frac{\delta{\cal L}_m}{\delta\phi}\right)\delta\phi\right]=\int d^4x\sqrt{-g}\left\{\frac{\phi}{2}\left[{\cal E}_{\mu\nu}-\frac{1}{\phi}T^{(m)}_{\mu\nu}\right]\delta g^{\mu\nu}\right.\nonumber\\
&\left.+\left[\frac{\omega}{\phi}\nabla^2\phi+\frac{1}{2}\left(\frac{\omega_{,\phi}}{\phi}-\frac{\omega}{\phi^2}\right)(\der\phi)^2+\frac{1}{2}R-V_{,\phi}+\frac{1}{2\phi}T^{(m)}\right]\delta\phi\right\}=0,\label{x}\end{align} where, in the last line, we have used the Ward identity \eqref{usef-rel} to replace $\delta{\cal L}_m/\delta\phi$ by $\sqrt{-g}\,T^{(m)}/2\phi$, and the following definition:

\begin{align} {\cal E}_{\mu\nu}:=G_{\mu\nu}-\frac{\omega}{\phi^2}\left[\der_\mu\phi\der_\nu\phi-\frac{1}{2}g_{\mu\nu}(\der\phi)^2\right]-\frac{1}{\phi}\left(\nabla_\mu\nabla_\nu-g_{\mu\nu}\nabla^2\right)\phi+g_{\mu\nu}\frac{V}{\phi}.\label{emn-def}\end{align} From equation \eqref{x}, by applying the variational principle of least action, the EBD and KG type EOMs:

\begin{align} \phi{\cal E}_{\mu\nu}&=T^{(m)}_{\mu\nu},\label{einst-eom}\\
2\omega\nabla^2\phi+\left(\omega_{,\phi}-\frac{\omega}{\phi}\right)(\der\phi)^2+\phi R&=2\phi V_{,\phi}-T^{(m)},\label{kg-eom}\end{align} are obtained. Note the difference between the KGBD-EOM \eqref{kg-eom} and \eqref{kgbd-eom'}, which is usually found in the bibliography.\footnote{Recall that \eqref{kgbd-eom'} is derived under the assumption that the matter Lagrangian density is not form-invariant under GCT \eqref{gen-gauge-t} and \eqref{mass-t}, which is not congruent with the transformation of masses $m\rightarrow\Omega^{-1}m$.} 

If we substitute the trace of EBD-EOM \eqref{einst-eom};

\begin{align} 3\nabla^2\phi+\frac{\omega}{\phi}(\der\phi)^2-\phi R=T^{(m)}-4V,\label{trace-eom}\end{align} in the KG-type equation \eqref{kg-eom}, we get the following EOM for the BD field: $(2\omega+3)\nabla^2\phi+\omega_{,\phi}(\der\phi)^2=2\phi V_{,\phi}-4V$, or, if conformal symmetry is taken into account, that is, $\phi V_{,\phi}=2V$ $\Rightarrow$ $V\propto\phi^2$, then the KG-type equation simplifies to the following:

\begin{align} (2\omega+3)\nabla^2\phi+\omega_{,\phi}(\der\phi)^2=0.\label{kg-eom'}\end{align} 

It is noticeable that this $\phi$-EOM does not depend on the curvature scalar or on the trace of the matter SET, such as, for example, the standard KG-type EOM \eqref{kgbd-eom'}. Hence, \eqref{kg-eom'} holds the same in the presence of matter and in a vacuum. That is, the $\phi$-EOM does not affect the gravitational dynamics. This property of the KG-type equation in conformal-form-invariant STG-JFBD theory is associated with gauge freedom arising from conformal symmetry. Actually, one is free to choose the coupling function. Once we choose a given $\omega=\omega(\phi)$, we replace it in \eqref{kg-eom'} and solve for $\phi$ to get a function $\phi=\phi(x)$. Otherwise, if we fix $\phi=\phi(x)$, and then substitute the chosen function $\phi(x)$ in \eqref{kg-eom'}, the latter can be integrated:

\begin{align} 2\omega+3=C\,e^{-2\int\frac{\nabla^2\phi}{(\der\phi)^2}\der_\mu\phi\,dx^\mu},\nonumber\end{align} where $C$ is an integration constant. From this equation we can determine $\omega=\omega(x)$. Hence, a choice of $\omega$ forces a choice of the BD field $\phi$ and vice versa. This is what we refer to as ``gauge choice'' in this theory. Note that in the particular case where $\omega=-3/2$, the KG-type equation \eqref{kg-eom'} becomes a vanishing identity $0\equiv 0$. In this case, the BD field does not obey an independent EOM, so a gauge choice amounts to choosing a specific function $\phi=\phi(x)$.



\subsection{Continuity equation and fifth force}


The divergence of the LHS of \eqref{einst-eom} yields

\begin{align} \nabla^\lambda\left(\phi\,{\cal E}_{\lambda\mu}\right)=-\frac{\der_\mu\phi}{2\phi}\left[2\omega\nabla^2\phi+\left(\omega_{,\phi}-\frac{\omega}{\phi}\right)(\der\phi)^2+\phi R-2\phi V_{,\phi}\right].\label{div-einst-eom}\end{align} Hence, if we substitute the KG-type equation \eqref{kg-eom} in \eqref{div-einst-eom}, taking into account that $\nabla^\lambda\left(\phi\,{\cal E}_{\lambda\mu}\right)=\nabla^\lambda T^{(m)}_{\lambda\mu}$, we get the following nonhomogeneous continuity equation:

\begin{align} \nabla^\lambda T^{(m)}_{\lambda\mu}=\frac{\der_\mu\phi}{2\phi}\,T^{(m)}.\label{nhom-cont-eq}\end{align} This means that a fifth force arises in conformal-invariant STG theory in the JFBD parametrization. The inhomogeneous term is required for conformal form-invariance to be a symmetry of the STG theory. 

Equation \eqref{nhom-cont-eq} is consistent with the EOM of timelike point-like particles; due to the assumption that the masses of timelike fields are point-dependent quantities $m_i(x)=\kappa_i\sqrt{\phi(x)}$ \cite{hobson-prd-2020, hobson-epjc-2022}, transforming as $m_i\rightarrow\Omega^{-1}m_i$ under GCT \cite{dicke-1962, bekenstein_1980}. The conformal form-invariant action for timelike point particles reads $S_i=\int m_i(x)ds=\kappa_i\int\sqrt{\phi(x)}ds$. Variation of this action leads to the EOM of point-like particles:

\begin{align} \frac{d^2x^\alpha}{ds^2}+\left\{^\alpha_{\mu\nu}\right\}\frac{dx^\mu}{ds}\frac{dx^\nu}{ds}=\frac{\der_\mu \phi}{2\phi}h^{\mu\alpha},\label{non-geod}\end{align} where the orthogonal projection tensor $h^{\mu\nu}$ is defined as

\begin{align} h^{\mu\alpha}:=g^{\mu\alpha}+u^\mu u^\alpha=g^{\mu\alpha}-\frac{dx^\mu}{ds}\frac{dx^\alpha}{ds}.\nonumber\end{align} The RHS of \eqref{non-geod} is required for conformal form-invariance to be a symmetry of this EOM. In contrast, for massless fields, the null geodesic equation of $V_4$ space:

\bea \frac{dk^\mu}{d\xi}+\left\{^\mu_{\nu\sigma}\right\}k^\nu k^\sigma=0,\label{0-geod}\eea where $k^\mu\equiv dx^\mu/d\xi$ is the wave vector ($k_\mu k^\mu=0$), and $\xi$ is an affine parameter along a null-geodesic, is conformally invariant as it is.

Note that the fifth force $f^\alpha=h^{\alpha\lambda}\der_\lambda\phi/2\phi$ is orthogonal to the fourth-velocity $u^\lambda f_\lambda=0$. Besides, it acts only on timelike fields, so it is a kind of ``dark force'', which does not interact with radiation and null-fields.



\section{Origin of the frame issue}
\label{sect-fiss}


Note the differences between standard STG-JFBD theory, where the frame issue arises, which is driven by the EOM \eqref{ebd-eom}, \eqref{kgbd-eom'}, and the conservation equation \eqref{jf-cons-eq}, and the present conformal form-invariant STG-JFBD theory represented by the EOM \eqref{einst-eom}, \eqref{kg-eom'}, and the nonhomogeneous continuity equation \eqref{nhom-cont-eq}. Both theories are given by the same overall action \eqref{tot-action}: $S_\text{tot}=\int d^4x\,{\cal L}_\text{tot}$, where, according to \eqref{tot-lag}:

\begin{align} {\cal L}_\text{tot}=\frac{\sqrt{-g}}{2}\left[\phi R-\frac{\omega}{\phi}(\der\phi)^2-2V\right]+{\cal L}_m(\chi,\der\chi,g_{\mu\nu}).\nonumber\end{align} However, depending on the conformal transformations considered and on the hypothesis about the variational derivative of the Lagrangian density of matter fields, the mathematical and physical consequences are different. 


To differentiate our conformal form-invariant approach to STG-JFBD theory from the standard one in the bibliography \cite{dicke-1962, morganstern_1970, anderson-1971, bekenstein_1980, cotsakis_1993, magnano_1994, capozziello_1997, kaloper_prd_1998, faraoni_rev, faraoni_ijmpd_1999, quiros_prd_2000, fujii_2007, faraoni_prd_2007, capozziello_prd_2013, quiros_grg_2013, morris-2014, sasaki_2016, quiros_ijmpd_2020, brans-dicke-1961, bergmann_1968, nordvedt_1970, ross_1972, fujii-1974, isenberg_1976, fujii_book, faraoni_book, quiros_ijmpd_2019}, let us expose the main hypotheses that underlie both approaches.


The standard approach to STG-JFBD theory is based on the following hypotheses: i) the symmetry transformations are restricted CTs \eqref{gauge-t}, ii) the masses of timelike fields transform as $m_i\rightarrow\Omega^{-1}m_i$ under CT \cite{dicke-1962, bekenstein_1980}, and iii) the Lagrangian density of matter is not conformal form-invariant, so the Ward identity \eqref{mat-ward-id} is not satisfied and, since ${\cal L}_m={\cal L}_m(\chi,\der\chi,g_{\mu\nu})$; that is, the Lagrangian density of matter fields $\chi$ does not depend explicitly on the BD field $\phi$, then $\delta{\cal L}_m/\delta\phi=0$.


The main hypotheses underlying our approach to conformal form-invariant STG-JFBD theory are: 

\begin{enumerate}

    \item The symmetry group is generated by form-invariance under GCT \eqref{gen-gauge-t},

    \item the masses of timelike fields are themselves point-dependent fields that transform as $m_i\rightarrow\Omega^{-1}m_i$ under CT; in particular, these can be written as \cite{hobson-prd-2020, hobson-epjc-2022}: $m_i(x)=\kappa_i\sqrt{\phi(x)}$,
    
\end{enumerate} The second hypothesis leads to the Lagrangian density of matter being conformal form-invariant, so the Ward identity \eqref{mat-ward-id} is satisfied:

\begin{align} \frac{\delta{\cal L}_m}{\delta\phi}=-\frac{1}{\phi}g^{\mu\nu}\frac{\delta{\cal L}_m}{\delta g^{\mu\nu}}=\frac{\sqrt{-g}}{2\phi}\,T^{(m)}\neq 0,\nonumber\end{align} which is related to the fact that ${\cal L}_m$ also depends on the BD field: ${\cal L}_m={\cal L}_m(\chi,\der\chi,g_{\mu\nu},\phi)$. Note that, in the standard approach in the bibliography, the second hypothesis above is not accompanied by fulfillment of the Ward identity, as it should be.

Equation \eqref{usef-rel} is central to obtaining the KG-type EOM \eqref{kg-eom'} and the nonhomogeneous continuity equation \eqref{nhom-cont-eq};

\begin{align} (2\omega+3)\nabla^2\phi+\omega_{,\phi}(\der\phi)^2=2\left(\phi V_{,\phi}-2V\right),\;\nabla^\lambda T^{(m)}_{\lambda\mu}=\frac{\der_\mu\phi}{2\phi}\,T^{(m)},\label{app-1}\end{align} respectively, which are form-invariant under GCT \eqref{gen-gauge-t}, instead of \eqref{kgbd-eom'} and the conservation equation \eqref{jf-cons-eq};

\begin{align} \left(3+2\omega\right)\nabla^2\phi+\omega_{,\phi}(\der\phi)^2=2\left(\phi V_{,\phi}-2V\right)+T^{(m)},\;\nabla^\lambda T^{(m)}_{\lambda\mu}=0,\label{app-2}\end{align} which are derived under the hypothesis that $\delta{\cal L}_m/\delta\phi=0$ and are not form-invariant under CT \eqref{gauge-t}, unless $T^{(m)}=0$. This is trivially true for the conservation equation (equation on the right in \eqref{app-2}). For the KG-type EOM (equation on the left in \eqref{app-2}), the demonstration requires some algebra. Let us, for simplicity of mathematical handling, rewrite the KG-type equation in terms of the coupling function $\sigma=\sigma(\phi)$ defined as;

\begin{align} \sigma(\phi):=\frac{3}{2}+\omega(\phi).\label{kappa}\end{align} We have

\begin{align} \nabla^2\phi+\frac{\sigma_{,\phi}}{2\sigma}(\der\phi)^2=\frac{\left(\phi V_{,\phi}-2V\right)}{\sigma}+\frac{T^{(m)}}{2\sigma}.\label{app-3}\end{align} Under the following GCT;

\begin{align} g_{\mu\nu}\rightarrow\Omega^2g_{\mu\nu},\;\phi\rightarrow\Omega^{-2}\phi,\;V\rightarrow\Omega^{-4}V,\;\sigma\rightarrow\frac{\sigma}{\left(1-2\frac{\Omega_{,\phi}}{\Omega}\phi\right)^2},\label{app-4}\end{align} the following transformation laws take place:

\begin{align} &\nabla^2\phi+\frac{\sigma_{,\phi}}{2\sigma}(\der\phi)^2\rightarrow\Omega^{-4}\left(1-2\frac{\Omega_{,\phi}}{\Omega}\phi\right)\left[\nabla^2\phi+\frac{\sigma_{,\phi}}{2\sigma}(\der\phi)^2\right],\nonumber\\
&\frac{\phi V_{,\phi}-2V}{\sigma}\rightarrow\Omega^{-4}\left(1-2\frac{\Omega_{,\phi}}{\Omega}\phi\right)\left(\frac{\phi V_{,\phi}-2V}{\sigma}\right),\;\frac{T^{(m)}}{2\sigma}\rightarrow\Omega^{-4}\left(1-2\frac{\Omega_{,\phi}}{\Omega}\phi\right)^2\frac{T^{(m)}}{2\sigma}.\nonumber\end{align} 

As seen from the transformation of the SET trace in the last line of the above equation (second term in the RHS of \eqref{app-3}), this term breaks the form-invariance of the KG-type equation under GCT \eqref{app-4}. Hence, it does not suffice to include the transformation of the coupling parameter $\omega$ (or $\sigma$) to allow for conformal form-invariance of the EOM of STG-JFBD theory. In addition to generalizing the conformal transformations as in \eqref{app-4}, it is required to remove the term with the trace of the matter SET $T^{(m)}$ from equation \eqref{app-3}. This is achieved precisely by taking into account the Ward identity \eqref{usef-rel}. This also implies that the nonhomogeneous continuity equation \eqref{nhom-cont-eq}, which is conformal form-invariant, takes place instead of the conservation equation \eqref{jf-cons-eq}, which is obviously not conformal form-invariant. This result confirms that the two different approaches to JFBD parametrization of STG theory differ in the consistent consideration of the second hypothesis above. In other words, the gravitational action of the STG-JFBD theory,

\begin{align} S_\text{grav}=\frac{1}{2}\int d^4x\sqrt{-g}\left[\phi R-\frac{\omega}{\phi}(\der\phi)^2-2V\right],\label{grav-action}\end{align} is form-invariant under GCT \eqref{gen-gauge-t} but is not conformal form-invariant under restricted CT \eqref{gauge-t}. Meanwhile, as shown in \cite{quiros-prd-2025}, the matter action, $S_m=\int d^4x\,{\cal L}_m(\chi,\der\chi,g_{\mu\nu})$, is conformal form-invariant as long as the mass parameter transforms as: $m\rightarrow\Omega^{-1}m$. This result passed unnoticed for many years and is one of the main reasons behind the CFI. The omission of the Ward identity \eqref{usef-rel} is another of the sources of the conformal frame issue. This identity is important for the correct derivation of the KG-type equation \eqref{kg-eom'}: $(2\omega+3)\nabla^2\phi+\omega_{,\phi}(\der\phi)^2=0$, instead of the well-known KG-JFBD equation \eqref{kgbd-eom'}: $\left(3+2\omega\right)\nabla^2\phi+\omega_{,\phi}(\der\phi)^2=2\left(\phi V_{,\phi}-2V\right)+T^{(m)}$. The Einstein-type equation \eqref{einst-eom} is the same in both cases: $\phi{\cal E}_{\mu\nu}=T^{(m)}_{\mu\nu}$.


\section{Passive and active conformal transformations}
\label{sect-act-pass}


Often, arguments about the physical equivalence of different conformal frames are biased by the passive approach to conformal transformations. This approach establishes that the different conformal sets of fields that give rise to conformal frames are physically irrelevant and that the physically meaningful quantities are necessarily conformal invariants \cite{jarv_2015}. As shown in \cite{quiros-prd-2025}, PACT is only one of the two possible approaches to conformal transformations. In this section, we discuss the physical implications of these two approaches: 1) PACT and 2) AACT, in STG theories. Here we give a more detailed and elaborated explanation of these approaches to conformal transformations than in \cite{quiros-prd-2025}, and we apply them to the investigation of the CFI.\footnote{We underline that the CT we will consider here: $g_{\mu\nu}\rightarrow\Omega^2g_{\mu\nu}$, no matter whether active or passive, does not imply coordinate transformations of either kind: active or passive. Active and passive CTs refer to transformations in phase space ${\cal M}_\text{fields}$; consequently, they do not belong to the conformal group of transformations $C(1,3)$. These can not be confounded with dilatation transformations, which imply coordinate transformations: $\delta x^\mu=\epsilon x^\mu$, nor with special CTs: $\delta x^\mu=2v_\nu x^\nu x^\mu-x^2v^\mu$, which are also coordinate transformations. Besides, here we consider that constants, regardless of whether they are fundamental or not, or whether they are dimensionless or dimensionful, are not transformed by conformal transformations.}

\begin{figure*}[t!]\centering
\includegraphics[width=8.5cm]{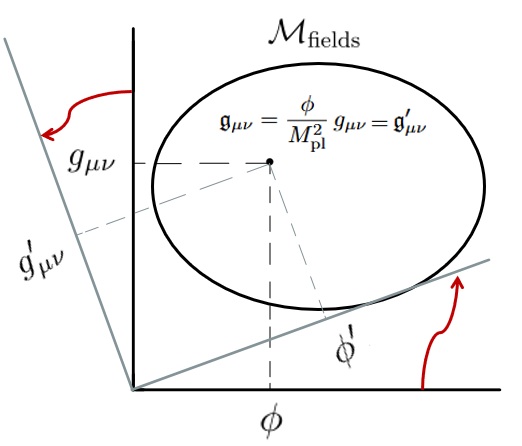}
\includegraphics[width=8.5cm]{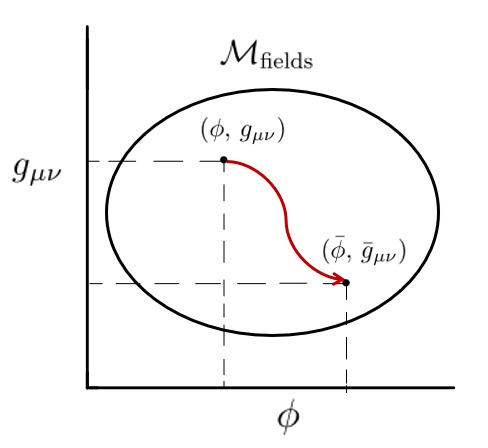}
\caption{Drawings of the configuration space manifold ${\cal M}_\text{fields}$ (oval region in the plane $\phi-g_{\mu\nu}$). Each point in the manifold represents a vacuum global gravitational state. In the left figure, the passive standpoint on the CT is illustrated. In this case the conformal transformation may be viewed as a ``rotation'' of the coordinate system $R:(\phi,g_{\mu\nu})$ in the plane $\phi-g_{\mu\nu},$ which leaves invariant the gravitational state $\mathfrak{G}_\mathfrak{g}:(\mathfrak{g}_{\mu\nu})$, where the invariant metric reads $\mathfrak{g}_{\mu\nu}=(\phi/M^2_\text{pl})\,g_{\mu\nu}$. According to the active point of view on the CT (right figure), the conformal transformation represents an actual ``motion'' of the point $(\phi,g_{\mu\nu})$ in ${\cal M}_\text{fields}$. That is, it represents a real change of the gravitational state $\mathfrak{G}_{\bf g}:(g_{\mu\nu},\phi)$ $\rightarrow{\bar{\mathfrak{G}}}_{\bf g}:(\bar g_{\mu\nu},\bar\phi)$.}\label{fig1}
\end{figure*}

In this document, we follow the geometric approach proposed in \cite{vilko-npb-1984} to search for a unique effective action in QFT and also undertaken in \cite{gong_2011} in the context of multifield inflation (see also \cite{stein-2013, jarv_2015, karamitsos_2016, karamitsos_2018, karamitsos_2020, karamitsos_2021}), where the scalar fields $\vphi_a$ ($a=1,2,...,N$) are treated as generalized coordinates living in the configuration space, so that any transformation of the scalar fields is then regarded as a coordinate transformation in the field space. We further push this approach and assume that not only the scalar field $\phi$, but also the metric tensor $g_{\mu\nu}$ and the matter fields $\chi$, are generalized coordinates in the configuration space manifold, which we represent by ${\cal M}_\text{fields}$. As shown in FIG. \ref{fig1} for the vacuum case, each point in ${\cal M}_\text{fields}$ represents a ``global gravitational state'' (GGS) of the system.\footnote{Here, by a global gravitational state we understand full knowledge of the fields at every spacetime point: $\phi=\phi(t,\vec{x})$, $g_{\mu\nu}=g_{\mu\nu}(t,\vec{x})$, $\chi_i=\chi_i(t,\vec{x})$ and $m=m(t,\vec{x})$; that is, each point in ${\cal M}_\text{fields}$ is a manifold itself.}


\subsection{Passive conformal transformations}


Passive CTs relate different representations of the same physical state, as illustrated in the left panel of FIG. \ref{fig1} for the vacuum case. Hence, the physically meaningful quantities must be invariant under the following conformal transformations:

\begin{align} g'_{\mu\nu}=\Omega'^2g_{\mu\nu},\;\phi'=\Omega'^{-2}\phi,\;\chi'=\Omega'^{w_\chi}\chi,\label{passive-t}\end{align} or, equivalently $g'_{\mu\nu}=(\phi/\phi')\,g_{\mu\nu}$, $\chi'=(\phi/\phi')^{w_\chi}\chi$. These transformations relate different representations: ${\cal R}_g:(g_{\mu\nu},\phi,\chi)$ and ${\cal R}'_g:(g'_{\mu\nu},\phi',\chi')$, of the same GGS $\mathfrak{G}_\mathfrak{g}:(\mathfrak{g}_{\mu\nu},\Psi)$ in ${\cal M}_\text{fields}$, where

\begin{align} \mathfrak{g}_{\mu\nu}=\frac{\phi}{M^2_\text{pl}}\,g_{\mu\nu},\;\Psi=\left(\frac{\phi}{M^2_\text{pl}}\right)^\frac{w_\chi}{2}\chi,\label{inv-met}\end{align} are the conformal invariant (physically meaningful) metric tensor, and the physically meaningful matter fields, respectively.\footnote{It should be noted that in the presence of matter fields $\chi$, in addition to \eqref{inv-met}, any conformal invariant combination such as $\mathfrak{g}_{\mu\nu}=\chi^{-2/w_\chi}g_{\mu\nu}$ can be used. However, when there is more than one matter field, a certain ambiguity arises because of the different ways in which a conformal invariant metric can be defined. In addition, mixing of gravitational and matter degrees of freedom (DOF) in an invariant metric tensor can lead to confusing results. For that reason and because the metric $g_{\mu\nu}$ and the scalar field $\phi$ are carriers of gravitational interactions in STG theories, here we take the combination $(\phi/M^2_\text{pl})\,g_{\mu\nu},$ to be the conformal invariant metric tensor. The scalar field $\phi$ can also be taken to construct conformal invariant matter fields \eqref{inv-met}.}


Under passive CT \eqref{passive-t}; $\mathfrak{g}'_{\mu\nu}=\phi'g'_{\mu\nu}=\phi\,g_{\mu\nu}=\mathfrak{g}_{\mu\nu}$, $\Psi'=\phi'^{\frac{w_\chi}{2}}\chi'=\phi^\frac{w_\chi}{2}\chi=\Psi$, where the metric $g_{\mu\nu}$, the scalar field $\phi$ and the matter fields $\chi$, are mere coordinates in ${\cal M}_\text{fields}$; that is, these are just auxiliary fields, the GGS $\mathfrak{G}_\mathfrak{g}:\left(\mathfrak{g}_{\mu\nu},\Psi\right)$ is conformal invariant, i.e. $\mathfrak{G}_\mathfrak{g}:\left(\mathfrak{g}_{\mu\nu},\Psi\right)=\mathfrak{G}'_\mathfrak{g}:\left(\mathfrak{g}'_{\mu\nu},\Psi'\right)$. Note that in the vacuum case, the gravitational state according to PACT is fully determined by the physical metric alone $\mathfrak{G}_\mathfrak{g}:\left(\mathfrak{g}_{\mu\nu}\right)$, as illustrated in the left panel of FIG. \ref{fig1}, while according to AACT it is given by the pair ${\cal S}_g:(g_{\mu\nu},\phi)$, as shown in the right panel of the figure.


In terms of the conformal invariant metric tensor, the line element reads:

\begin{align} d\mathfrak{s}^2=\mathfrak{g}_{\mu\nu}dx^\mu dx^\nu,\label{triv-le}\end{align} while the physical curvature invariants, such as the curvature scalar and the Kretschmann invariant of the physical metric, are given by the following expressions; $\mathfrak{R}=\mathfrak{g}^{\mu\nu}\mathfrak{R}_{\mu\nu}$, $K(\mathfrak{g}):=\mathfrak{R}^{\sigma\mu\lambda\nu}\mathfrak{R}_{\sigma\mu\lambda\nu}$, where $\mathfrak{R}_{\mu\nu}=\mathfrak{g}^{\lambda\sigma}\mathfrak{R}_{\lambda\mu\sigma\nu}$ is the Ricci tensor and $\mathfrak{R}^\alpha_{\;\;\mu\beta\nu}$ is the Riemann-Christoffel curvature tensor of the physical metric $\mathfrak{g}_{\mu\nu}.$ These quantities are defined with respect to the affine connection: 

\begin{align}\mathfrak{C}^\alpha_{\;\;\mu\nu}:=\frac{1}{2}\mathfrak{g}^{\alpha\lambda}\left(\der_\nu\mathfrak{g}_{\mu\lambda}+\der_\mu\mathfrak{g}_{\nu\lambda}-\der_\lambda\mathfrak{g}_{\mu\nu}\right),\label{triv-lc-c}\end{align} which coincides with the Levi-Civita (LC) connection of the physical metric. 


We shall demonstrate that passive CT \eqref{passive-t} cannot be an actual symmetry of theory because it acts on the auxiliary fields $g_{\mu\nu}$, $\phi$, and $\chi$, which are removed from the covariant description in ${\cal M}_\text{fields}$. Actually, in a covariant formulation of STG theory in the configuration space, all of the auxiliary fields may be safely removed, so that the main equations of the theory can be written in terms of the conformal invariant (physical) quantities alone. This is to be contrasted, for example, with form-invariance of the EOM under general coordinate transformations since the coordinates cannot be removed from the mathematical description of the theory, given that these are in the derivatives of the fields, etc.


Written in terms of the conformal invariant (physical) quantities, the conformal transformations \eqref{passive-t} become the identity transformation: $\mathfrak{g}_{\mu\nu}\rightarrow\mathfrak{g}_{\mu\nu}$, $\sqrt{-\mathfrak{g}}\rightarrow\sqrt{-\mathfrak{g}}$, $\mathfrak{C}^\alpha_{\;\;\mu\nu}\rightarrow\mathfrak{C}^\alpha_{\;\;\mu\nu}$, $\mathfrak{R}^\alpha_{\;\;\mu\beta\nu}\rightarrow\mathfrak{R}^\alpha_{\;\;\mu\beta\nu}$, $\mathfrak{R}_{\mu\nu}\rightarrow\mathfrak{R}_{\mu\nu}$, $\mathfrak{R}\rightarrow\mathfrak{R}$; that is, a useless trivialization of CT. Moreover, any simultaneous transformations of the auxiliary fields that leave the physical fields invariant, such as passive CT \eqref{passive-t}, are obviously a fictitious symmetry of the theory as long as it is written in terms of the physical quantities. This is in line with the result discussed in \cite{woodard-1986} and later demonstrated in \cite{jackiw-2015}, based on the Noether symmetry approach (see also \cite{oda-2022, rodrigo-arxiv},) that the Noether current corresponding to conformal symmetry is identically vanishing, so that passive CT \eqref{passive-t} is a ``fake transformation.''



\subsection{Active conformal transformations}


In contrast, active CT:

\begin{align} \bar g_{\mu\nu}=\bar\Omega^2g_{\mu\nu},\;\bar\phi=\bar\Omega^{-2}\phi,\;\bar\chi=\bar\Omega^{w_\chi}\chi,\label{active-t}\end{align} relate different physical fields $g_{\mu\nu}\Leftrightarrow\bar g_{\mu\nu}$, $\phi\Leftrightarrow\bar\phi$, and $\chi\Leftrightarrow\bar\chi$. That is, these relate different global gravitational states (GGSs) $\mathfrak{G}_g:(g_{\mu\nu},\phi,\chi)$ and $\bar{\mathfrak{G}}_g:(\bar g_{\mu\nu},\bar\phi,\bar\chi)$ in the configuration space ${\cal M}_\text{fields}$, as illustrated in the right panel of FIG. \ref{fig1} for the vacuum case.\footnote{By a global gravitational state, we mean the full knowledge of the fields at every spacetime point: $g_{\mu\nu}=g_{\mu\nu}(t,\vec{x})$, $\phi=\phi(t,\vec{x})$, and $\chi=\chi(t,\vec{x})$.} The different GGSs are characterized, for example, by different curvature invariants. In this case, the physically meaningful quantities are not necessarily conformal invariant quantities. In particular, the metric $g_{\mu\nu}$, the scalar field $\phi$, and the matter fields $\chi$ have themselves independent physical meaning. In this case, the conformal invariants like $\mathfrak{g}_{\mu\nu}$ and $\Psi$:

\begin{align} \mathfrak{g}_{\mu\nu}=\frac{\phi}{M^2_\text{pl}}\,g_{\mu\nu}=\frac{\bar\phi}{M^2_\text{pl}}\,\bar g_{\mu\nu},\;\Psi=\left(\frac{\phi}{M^2_\text{pl}}\right)^\frac{w_\chi}{2}\chi=\left(\frac{\bar\phi}{M^2_\text{pl}}\right)^\frac{w_{\bar\chi}}{2}\bar\chi,\label{inv-rel}\end{align} are useful in establishing the relationships between the physical fields $g_{\mu\nu}$, $\phi$ and $\chi$, in different gauges:

\begin{align} \bar g_{\mu\nu}=\left(\frac{\phi}{\bar\phi}\right)g_{\mu\nu},\;\bar\chi=\left(\frac{\phi}{\bar\phi}\right)^\frac{w_\chi}{2}\chi.\label{g-active}\end{align}



\subsection{Matter fields}


It has been demonstrated in \cite{quiros-prd-2025} for fundamental matter fields as well as for perfect fluids that, if the mass parameter is point-dependent $m=m(x)$, such that under CT \eqref{gauge-t} $m(x)\rightarrow\Omega^{-1}m(x)$ and the energy density of perfect fluids transforms as $\rho(x)\rightarrow\Omega^{-4}\rho(x)$, the standard Lagrangian density of matter is conformal form-invariant, as seen in \eqref{lag-t}: ${\cal L}_m(\chi,\der\chi,g_{\mu\nu})\rightarrow{\cal L}_m(\hat\chi,\hat\der\hat\chi,\hat g_{\mu\nu})={\cal L}_m(\chi,\der\chi,g_{\mu\nu})$. It is not difficult to show that the above Lagrangian density is equivalent to the same Lagrangian density written in terms of conformal invariant quantities:

\begin{align} {\cal L}_m(\chi,\der\chi,g_{\mu\nu})={\cal L}_m(\Psi,\mathfrak{d}\Psi,\mathfrak{g}_{\mu\nu}),\label{mat-lag-equal}\end{align} where the conformal invariant metric $\mathfrak{g}_{\mu\nu}$ and the conformal invariant matter fields $\Psi$ are given by \eqref{inv-met}. For example, for the Lagrangian density of the Dirac fermion, we have

\begin{align} \sqrt{-g}\,\bar\psi\left(i\cancel{\cal D}+m\right)\psi=\sqrt{-\mathfrak{g}}\,\bar\Psi\left(i\cancel{\mathfrak{D}}+\mathfrak{m}\right)\Psi,\label{dirac-equal}\end{align} where the gauge covariant derivative $\mathfrak{D}_\mu$ is defined in the same way as ${\cal D}_\mu$, but with the following replacements (we use the notation of \cite{quiros-prd-2025}, see, in particular, Section III.B): $g_{\mu\nu}\rightarrow\mathfrak{g}_{\mu\nu}=(\phi/M^2_\text{pl})\,g_{\mu\nu}$, $e^a_\mu\rightarrow\mathfrak{e}^a_\mu=(\sqrt\phi/M_\text{pl})\,e^a_\mu$, etc. In \eqref{dirac-equal}, the conformal invariant fermion spinor and mass

\begin{align} \Psi=\left(\frac{\sqrt\phi}{M_\text{pl}}\right)^{-\frac{3}{2}}\psi,\;\mathfrak{m}=\left(\frac{M_\text{pl}}{\sqrt\phi}\right)\,m,\nonumber\end{align} have been introduced. Equation \eqref{dirac-equal} implies that ${\cal L}_m(\psi,{\cal D}\psi,g_{\mu\nu})={\cal L}_m(\Psi,\mathfrak{D}\Psi,\mathfrak{g}_{\mu\nu})$, thus confirming \eqref{mat-lag-equal}.


The equality \eqref{mat-lag-equal} also takes place for the Lagrangian density of the prefect fluid. Actually, for a perfect fluid with energy density $\rho$ and barotropic pressure $p=w\rho$, where the constant $w$ is the equation of state (EOS) parameter, the Lagrangian density can be written in the following form \cite{hawking-book, faraoni_2009, berto-prd-2008}:\footnote{Although in the bibliography one also finds that ${\cal L}_\text{fluid}=\sqrt{-g}\,p$ (see, for instance, \cite{schutz_1970}), unless the perfect fluid couples explicitly to the curvature, which is not the case in this paper, both Lagrangian densities are equivalent \cite{faraoni_2009}.}

\begin{align} {\cal L}_\text{fluid}(\rho,g_{\mu\nu})=-\sqrt{-g}\,\rho.\label{pfluid-lag}\end{align} Let us define the following conformal invariant energy density of matter: $\mathfrak{r}:=(M^2_\text{pl}/\phi)^2\,\rho$. The Lagrangian density of matter fields can be written as

\begin{align} {\cal L}_\text{fluid}(\mathfrak{r},\mathfrak{g}_{\mu\nu})=-\sqrt{-\mathfrak{g}}\,\mathfrak{r}.\label{inv-m-lag}\end{align} 

If we take into account the definition \eqref{inv-met} of the conformal invariant metric, that is, $\sqrt{-\mathfrak{g}}=(\phi/M^2_\text{pl})^2\sqrt{-g}$. Then, it is evident that $\sqrt{-\mathfrak{g}}\,\mathfrak{r}=\sqrt{-g}\,\rho$; so, the following equality takes place:

\begin{align} {\cal L}_\text{fluid}(\rho,g_{\mu\nu})={\cal L}_\text{fluid}(\mathfrak{r},\mathfrak{g}_{\mu\nu}),\label{equiv}\end{align} thus confirming \eqref{mat-lag-equal}. The left-hand side (LHS) of this equivalence is form-invariant under simultaneous CT \eqref{gauge-t} and $\rho\rightarrow\hat\rho=\Omega^{-4}\rho$, while the fields in its RHS, being conformal invariant quantities, are not transformed by CT. In other words: in the LHS of equivalence \eqref{mat-lag-equal}, the Lagrangian density of matter is written in terms of the fields that suffer conformal transformations \eqref{gauge-t}, while, in the RHS, the Lagrangian density of matter is written in terms of conformal invariant quantities. The equivalence implies that both the metric tensors $g_{\mu\nu}$ and $\mathfrak{g}_{\mu\nu}$ can be considered physical fields, since in both cases the matter fields are minimally coupled to the corresponding metric tensor. 

However, the main importance of the equivalence \eqref{mat-lag-equal} is that it serves as a mathematical foundation for the separation of CT into passive and active conformal transformations. The use of one form or another of the Lagrangian density of matter: LHS or RHS of \eqref{mat-lag-equal}, depends on the approach chosen. If we choose the PACT, then the Lagrangian density of matter on the RHS is the correct one. Otherwise, if one assumes the AACT, then the matter Lagrangian density in the LHS must be chosen.


From the LHS of \eqref{equiv}, it can be observed that, since under the active conformal transformations \eqref{active-t}:

\begin{align} \bar g_{\mu\nu}=\bar\Omega^2g_{\mu\nu},\;\bar\phi=\bar\Omega^{-2}\phi,\;\bar\rho=\bar\Omega^{-4}\rho,\nonumber\end{align} the Lagrangian density of matter is conformal form-invariant: ${\cal L}_\text{fluid}(\rho,g_{\mu\nu})={\cal L}_\text{fluid}(\bar\rho,\bar g_{\mu\nu})$, the conformal metric $\bar g_{\mu\nu}$ is the physical metric in the conformal frame as long as $g_{\mu\nu}$ is the physical metric in the original frame. This is a demonstration that, if we adopt the AACT, any metric that is conformal to some physical metric tensor is itself physical. That is, minimal coupling of matter is not modified by the conformal transformation according to the active point of view on CT.



\subsection{Remark on passive and active approaches}


Some readers might think that the active approach to conformal transformations (right figure in FIG. \ref{fig1}) is merely a mathematical or philosophical artifice without physical consequences. Here, we will attempt to convince them that the AACT actually embodies new physics.

As the starting point of our line of reasoning, let us agree on which metric to consider as the ``physical metric''. Usually, this is defined as the metric tensor to which the matter fields are minimally coupled. Since clocks are made out of these fields, this metric tensor defines the proper time measured by the clocks. Hence, the metric to which the matter fields are minimally coupled is the metric that has operational meaning \cite{brans-1988}, that is, the ``physical metric''. For example, if we look at the Lagrangian density of matter ${\cal L}_m={\cal L}_m(\chi,\der\chi,g_{\mu\nu})$, it is inferred that the matter fields $\chi$ are minimally coupled to the metric $g_{\mu\nu}$, so this is the physical metric; that is, the metric tensor that defines the proper time measured by clocks. However, if we look at the Lagrangian density ${\cal L}_m={\cal L}_m(\Psi,\mathfrak{d}\Psi,\mathfrak{g}_{\mu\nu})$, where the conformal invariant metric tensor $\mathfrak{g}_{\mu\nu}$ and the conformal invariant matter fields $\Psi$ are defined in \eqref{inv-met}, we conclude that the physical metric coincides with $\mathfrak{g}_{\mu\nu}$, since the conformal invariant matter fields $\Psi$ are minimally coupled to this metric tensor.\footnote{Note that the above Lagrangian density can be written alternatively as; ${\cal L}_m={\cal L}_m(\Psi,\mathfrak{d}\Psi,\phi\,g_{\mu\nu})$, where we explicitly write the definition of the conformal invariant metric tensor \eqref{inv-met} (for simplicity, we assume $M^2_\text{pl}=1$). From this form of the Lagrangian density ${\cal L}_m$, it follows that the physical metric is the conformal invariant product $\phi\,g_{\mu\nu}$, instead of the auxiliary metric $g_{\mu\nu}$.} Once the physical metric is identified in each case, we need to take into account the equation \eqref{mat-lag-equal}: ${\cal L}_m(\chi,\der\chi,g_{\mu\nu})\equiv{\cal L}_m(\Psi,\mathfrak{d}\Psi,\mathfrak{g}_{\mu\nu})$. 

For illustration, let us consider the following conformal form-invariant total Lagrangian density:

\begin{align} {\cal L}_\text{tot}=\frac{\sqrt{-g}}{2}\left[\phi R+\frac{3}{2\phi}(\der\phi)^2-2V_0\,\phi^2\right]+{\cal L}_m,\label{tot-lag-gen}\end{align} which is a particular member of the class of conformal form-invariant STG theory where the coupling parameter is $\omega=-3/2$. Note that in \eqref{tot-lag-gen} we can substitute any of the two equivalent matter Lagrangians: ${\cal L}_m(\chi,\der\chi,g_{\mu\nu})$ or ${\cal L}_m(\Psi,\mathfrak{d}\Psi,\mathfrak{g}_{\mu\nu})$. If we substitute the latter Lagrangian density, we get that,

\begin{align} {\cal L}_\text{tot}=\frac{\sqrt{-g}}{2}\left[\phi R+\frac{3}{2\phi}(\der\phi)^2-2V_0\,\phi^2\right]+{\cal L}_m(\Psi,\mathfrak{d}\Psi,\mathfrak{g}_{\mu\nu}).\nonumber\end{align} The next step is to rewrite the gravitational action in terms of the physical metric $\mathfrak{g}_{\mu\nu}$. This results in the Einstein-Hilbert Lagrangian density coupled to the matter fields:

\begin{align} {\cal L}_\text{tot}=\frac{\sqrt{-\mathfrak{g}}}{2}\,M^2_\text{pl}\left(\mathfrak{R}-2V_0\right)+{\cal L}_m(\Psi,\mathfrak{d}\Psi,\mathfrak{g}_{\mu\nu}),\label{eh-lag}\end{align} where $\mathfrak{R}$ is the curvature scalar of the physical metric $\mathfrak{g}_{\mu\nu}$. The Lagrangian density \eqref{eh-lag} is the covariant expression of the Lagrangian density \eqref{tot-lag-gen} in the configuration space. It is the result of applying the passive approach to conformal transformations acting on \eqref{tot-lag-gen}. According to PACT, the physically meaningful (measured) quantities are conformal invariants; that is, the metric $\mathfrak{g}_{\mu\nu}$, and the matter fields $\Psi$. 

As we can check by applying the conformal transformations to the physical fields: $\mathfrak{g}_{\mu\nu}\rightarrow\Omega^2\mathfrak{g}_{\mu\nu}$, $\Psi\rightarrow\Omega^{w_\Psi}\Psi$, conformal invariance is not a symmetry of \eqref{eh-lag}. Note that the auxiliary fields $g_{\mu\nu}$, $\phi$, and $\chi$, on which act the conformal transformations \eqref{gauge-t}, are removed from the covariant description \eqref{eh-lag}. For that reason, we say that PACT is not a suitable approach to investigate conformal symmetry.

If we substitute the left-hand matter Lagrangian density from \eqref{mat-lag-equal} in \eqref{tot-lag-gen}, we get the conformal form-invariant Lagrangian density above according to AACT, which is given by;

\begin{align} {\cal L}_\text{tot}=\frac{\sqrt{-g}}{2}\left[\phi R+\frac{3}{2\phi}(\der\phi)^2-2V_0\,\phi^2\right]+{\cal L}_m(\chi,\der\chi,g_{\mu\nu}).\label{aact-lag}\end{align} In this case, $g_{\mu\nu}$, which is an auxiliary field according to PACT, is the physical metric according to the active approach, so we do not need to rewrite the gravitational Lagrangian density in terms of some other metric, as we did above. The Lagrangian densities \eqref{eh-lag} and \eqref{aact-lag} are fully equivalent. However, the physical consequences are drastically different. 

For example, conformal invariance is not a symmetry of \eqref{eh-lag}, while it is a symmetry of \eqref{aact-lag}. Besides, according to PACT, only the GR phenomenology based on \eqref{eh-lag} makes physical sense, while, according to AACT, the phenomenology associated with the metric $g_{\mu\nu}$, as well as the phenomenology associated with any conformal metric $\hat g_{\mu\nu}=\Omega^2g_{\mu\nu}$, make physical sense. Equation \eqref{lag-t}; ${\cal L}_m(\hat\chi,\hat\der\hat\chi,\hat g_{\mu\nu})={\cal L}_m(\chi,\der\chi,g_{\mu\nu})$, expresses the form-invariance of the matter Lagrangian density under the conformal transformations \cite{quiros-prd-2025}: $g_{\mu\nu}=\Omega^2g_{\mu\nu}$, $\hat\phi=\Omega^{-2}\phi$, $\hat\chi=\Omega^{w_\chi}\chi$. Since matter fields $\chi$ are minimally coupled to the metric $g_{\mu\nu}$, this is the physical metric in this gauge, while conformal matter fields $\hat\chi$ are minimally coupled to the conformal metric $\hat g_{\mu\nu}$, so the latter is the physical metric in the conformal gauge. Being both physical (and different), both lead to different phenomenology, which must be tested experimentally. 

In addition, according to PACT, the covariant Lagrangian \eqref{eh-lag} describes the standard phenomenology of general relativity in Riemann space $V_4$. This means, for example, that the masses of point particles are constant parameters. In contrast, since according to AACT the physically meaningful Lagrangian density is given by \eqref{aact-lag}, the masses of timelike fields must be point-dependent quantities for conformal invariance to be a symmetry of STG theory \cite{dicke-1962, bekenstein_1980, casas_1992}. Although this mass property cannot be revealed in local experiments since the measuring sticks are made of fields whose masses vary in the same universal way as the other fields in nature, in experiments implying exchange of information between distant points in spacetime, as, for example, the redshift measurements, this property may be evaluated.



\section{STG-JFBD theory in the Configuration Space}
\label{sect-conf-space}


We already made clear that passive CT relates different representations of the same global gravitational state (see the left panel of FIG. \ref{fig1}). In this case, the conformal invariants are the physically significant quantities of the theory. In particular, the conformal invariant quantities:\footnote{From now on, for simplicity of writing, we use the units system where $M_\text{pl}=1$. To go back to the standard units, we must replace everywhere the scalar field $\phi$ by $\phi/M^2_\text{pl}$.} $\mathfrak{g}_{\mu\nu}=\phi\,g_{\mu\nu}=\hat\phi\,\hat g_{\mu\nu}=\hat{\mathfrak{g}}_{\mu\nu}$ and $\Psi=\phi^\frac{w_\chi}{2}\chi=\hat\phi^\frac{w_\chi}{2}\hat\chi=\hat\Psi,$ are the physical metric and matter fields, respectively. The auxiliary fields (those that are acted on by the CT): $\chi$, $\phi$, and $g_{\mu\nu}$, are just coordinates in the configuration space ${\cal M}_\text{fields}$.

Our goal now is to write the overall Lagrangian density \eqref{tot-lag} of JFBD-STG theory, in terms of the conformal invariant (physical) metric alone, so STG theory in JFBD parametrization will be written covariantly in the configuration space ${\cal M}_\text{fields}$. The total Lagrangian density is given by

\begin{align} {\cal L}_\text{tot}=\frac{\sqrt{-g}}{2}\left[\phi R-\frac{\omega}{\phi}(\der\phi)^2-2V\right]+{\cal L}_m(\Psi,\mathfrak{d}\Psi,\mathfrak{g}_{\mu\nu}),\label{tot-lag-pact}\end{align} where, due to the equivalence \eqref{mat-lag-equal}, the matter Lagrangian density is written in terms of the conformal invariant fields $\mathfrak{g}_{\mu\nu}$, $\Psi$, and its derivatives. Under CT, this Lagrangian is not modified, since conformal transformations of the auxiliary fields (the coordinates in configuration space) amount to the identity transformation of the conformal invariant fields.


Here, we shall prove that if we enlarge the conformal transformations to include the transformation of the coupling function and of the mass parameter, that is, if we consider the following GCT 

\begin{align} g_{\mu\nu}\rightarrow\hat g_{\mu\nu}=\Omega^2g_{\mu\nu},\;\phi\rightarrow\hat\phi=\Omega^{-2}\phi,\;V\rightarrow\hat V=\Omega^{-4}V,\;\sigma\rightarrow\hat\sigma=\frac{\sigma}{\left(1-2\frac{\Omega_{,\phi}}{\Omega}\phi\right)^2},\;m\to\hat m=\Omega^{-1}m,\label{gen-gauge-t'}\end{align} where, for compactness of writing, we have introduced the redefinition \eqref{kappa}: $\sigma(\phi)\equiv\omega(\phi)+3/2$, the STG theory in JFBD parametrization \eqref{tot-lag}, admits a covariant formulation under PACT even in the presence of matter fields.\footnote{The transformation \eqref{mass-t} of the mass parameter is included here, precisely, to warrant that the matter Lagrangian density is form-invariant under the resulting GCTs.} The latter case generalizes the framework investigated in \cite{faraoni_1998}, where only the vacuum case was considered. 


It has been shown above that the gravitational Lagrangian density of STG theory in JFBD parametrization \eqref{jfbd-lag}, can be rewritten in terms of physical quantities plus the kinetic term of the auxiliary scalar field $\phi$. The total Lagrangian density \eqref{tot-lag-pact} can be written as

\begin{align} {\cal L}_\text{tot}=\frac{\sqrt{-\mathfrak{g}}}{2}\left\{\mathfrak{R}-\sigma(\phi)\frac{(\mathfrak{D}\phi)^2}{\phi^2}-2{\cal V}\right\}+{\cal L}_m(\Psi,\mathfrak{d}\Psi,\mathfrak{g}_{\mu\nu}).\label{pact-tot-lag}\end{align} The above gravitational Lagrangian density is not form-invariant under CT \eqref{gauge-t}. However, as long as the transformation of the coupling function is applied simultaneously with \eqref{gauge-t}, the total Lagrangian density \eqref{pact-tot-lag} becomes form-invariant under GCT \eqref{gen-gauge-t'}. Let us demonstrate this result. Under the transformation of $\omega=\omega(\phi)$ \eqref{w-gauge-t}, the parameter $\sigma=\sigma(\phi)$ transforms as

\begin{align} \sigma\rightarrow\hat\sigma=\frac{\sigma}{\left(1-2\frac{\Omega_{,\phi}}{\Omega}\phi\right)^2},\label{sigma-t}\end{align} while, under \eqref{gauge-t};

\begin{align} \frac{(\mathfrak{D}\phi)^2}{\phi^2}\rightarrow\left(1-2\frac{\Omega_{,\phi}}{\Omega}\phi\right)^2\frac{(\mathfrak{D}\phi)^2}{\phi^2}.\nonumber\end{align} Hence, the product $\sigma(\phi)(\mathfrak{D}\phi)^2/\phi^2$ remains form-invariant and, consequently, the total Lagrangian density ${\cal L}_\text{tot}$ is form-invariant, under GCT \eqref{gen-gauge-t'}.


It makes sense to carry out the following scalar field redefinition:

\begin{align} \Phi=\int\frac{\sqrt{\sigma(\phi)}}{\phi}\,d\phi,\label{Phi}\end{align} where the scalar field $\Phi$ is conformal invariant. As a consequence, \eqref{pact-tot-lag} can be written as follows:

\begin{align} {\cal L}_\text{tot}=\frac{\sqrt{-\mathfrak{g}}}{2}\left\{\mathfrak{R}-(\mathfrak{D}\Phi)^2-2{\cal V}\right\}+{\cal L}_m(\Psi,\mathfrak{d}\Psi,\mathfrak{g}_{\mu\nu}),\label{pact-tot-lag'}\end{align} which is the GR Lagrangian density with background matter fields $\Phi$ and $\Psi$. Notice that since $V=V_0\,\phi^2$ is the only potential that is compatible with the transformation property $V\rightarrow\Omega^{-4}V$, ${\cal V}=V_0$ is actually a constant. In \eqref{pact-tot-lag'} only conformal invariant quantities appear, so the overall Lagrangian density of JFBD-STG theory is written covariantly in the configuration space.\footnote{The overall Lagrangian density ${\cal L}_\text{tot}$ is trivially conformal form-invariant since the conformal transformations act on the auxiliary fields $g_{\mu\nu}$, $\phi$, $\chi$, as well as on the field-dependent parameters $V=V(\phi)$ and $\omega=\omega(\phi)$, but not on the physical fields $\mathfrak{g}_{\mu\nu}$, $\Phi$, and $\Psi$. Once the auxiliary fields are removed from the Lagrangian density, and consequently, from the EOM of the STG theory, the conformal transformations become the identity transformation of the physical fields. That is, the conformal transformation amounts to the identity transformation, so conformal invariance can be, at most, a fictitious symmetry.} A similar result has been previously found in \cite{jarv_2015} in a bit different way.


Taking the conformal invariant divergence of the Einstein EOM derived from \eqref{pact-tot-lag'}, and bearing in mind the KG-EOM $\mathfrak{D}^2\Phi={\cal V}_{,\Phi}$ and the Bianchi identity $\mathfrak{D}^\lambda\mathfrak{G}_{\lambda\mu}=0$, where $\mathfrak{G}_{\lambda\mu}$ is the Einstein tensor in terms of the conformal invariant metric, we obtain the standard conservation equation:

\begin{align} \mathfrak{D}^\lambda{\cal T}^{(m)}_{\lambda\mu}=0,\label{div-set-cinv'}\end{align} written in covariant form. 

The discussion in this section concludes that, according to the passive approach to conformal transformations, the conformal invariant JFBD-STG theory \eqref{tot-lag-pact} does not carry any new phenomenology beyond GR theory with (minimally coupled) scalar field matter \eqref{pact-tot-lag'}. Fortunately, there is another possible perspective on CTs: the active approach to conformal transformations.



\section{Gauge freedom in the present framework}
\label{sect-gauges}


The framework we develop in this document is designed to identify the consequences of conformal symmetry in STG theory within the JFBD parametrization. It is based on the following statements:

\begin{enumerate}
    
    \item The JFBD-STG theory \eqref{tot-lag} is form-invariant under generalized conformal transformations \eqref{gen-gauge-t}, \eqref{mass-t}.

    \item Since the mass parameter transforms according to \eqref{mass-t}, then the Lagrangian density of matter fields is conformal form-invariant. Consequently, the Ward identity \eqref{usef-rel} holds.

    \item The active approach to conformal transformations is the only possible understanding of CTs, where these carry phenomenological consequences. Here, we follow this approach.
    
\end{enumerate}

According to AACT, the quantities in the overall Lagrangian density \eqref{tot-lag} (here we assume the conformal form-invariant potential $V=\lambda\phi^2$, where $\lambda$ is a constant parameter):

\begin{align} {\cal L}_\text{tot}=\frac{\sqrt{-g}}{2}\left[\phi R-\frac{\sigma-3/2}{\phi}(\der\phi)^2-2\lambda\phi^2\right]+{\cal L}_m(\chi,\der\chi,g_{\mu\nu}),\label{aact-tot-lag}\end{align} of JFBD-STG theory, in particular the fields $g_{\mu\nu}$, $\phi$ and $\chi$, as they are, are physical (measurable) quantities. This is explicit in the form of the matter Lagrangian density, where it is seen that the matter fields $\chi$ are minimally coupled to the metric $g_{\mu\nu}$. This means that the metric tensor $g_{\mu\nu}$ defines proper-time measurements. Under a generalized conformal transformation \eqref{gen-gauge-t'} plus $\chi\to\hat\chi=\Omega^{w_\chi}\chi$, or, if we substitute $\Omega=\sqrt{\phi/\hat\phi}$:

\begin{align} \hat g_{\mu\nu}=\left(\frac{\phi}{\hat\phi}\right)g_{\mu\nu},\;\hat\chi=\left(\frac{\phi}{\hat\phi}\right)^\frac{w_\chi}{2}\chi,\;\hat\sigma=\frac{1}{(\hat\phi_{,\phi})^2}\left(\frac{\hat\phi}{\phi}\right)^2\sigma,\;\hat m=\sqrt\frac{\hat\phi}{\phi}\,m,\label{aact-gct}\end{align} we get

\begin{align} {\cal L}_\text{tot}=\frac{\sqrt{-\hat g}}{2}\left[\hat\phi\hat R-\frac{\hat\sigma-3/2}{\hat\phi}(\hat\der\hat\phi)^2-2\lambda\hat\phi^2\right]+{\cal L}_m(\hat\chi,\hat\der\hat\chi,\hat g_{\mu\nu}).\label{aact-tot-lag'}\end{align} It is evident that the overall Lagrangian density \eqref{aact-tot-lag} is form-invariant under \eqref{aact-gct}. Moreover, it is also clear that in the conformal gauge, the conformal matter fields $\hat\chi$ are minimally coupled to the conformal metric $\hat g_{\mu\nu}$, which is the physical metric in this gauge. Hence, both conformally related metric tensors are physical metrics in distinct gauges. Each yields a different phenomenology associated with different global gravitational states: $\mathfrak{G}:(\phi,g_{\mu\nu},\chi,m)$ and $\hat{\mathfrak{G}}:(\hat\phi,\hat g_{\mu\nu},\hat\chi,\hat m)$. Both GGSs belong to the same conformal class. In general, any two different gauges $\mathfrak{G}_i:(\phi_i,g^{(i)}_{\mu\nu},\chi_i,m_i)$ and $\mathfrak{G}_j:(\phi_j,g^{(j)}_{\mu\nu},\chi_j,m_j)$, where $i,j=0,1,2,...,N$ ($N\to\infty$), that belong to the same conformal equivalence class $\mathfrak{C}:(\mathfrak{G}_0,\mathfrak{G}_1,...,\mathfrak{G}_i,...,\mathfrak{G}_N)$, are related by the following equations:

\begin{align} g^{(i)}_{\mu\nu}=\left(\frac{\phi_j}{\phi_i}\right)g^{(j)}_{\mu\nu},\;\hat\chi_i=\left(\frac{\phi_j}{\phi_i}\right)^\frac{w_\chi}{2}\chi_j,\;\sigma_i=\frac{1}{(\phi_{i,\phi_j})^2}\left(\frac{\phi_i}{\phi_j}\right)^2\sigma_j,\;m_i=\sqrt\frac{\phi_i}{\phi_j}\,m_j,\nonumber\end{align} which are residuals of conformal symmetry.\footnote{Similar to the loss of explicit coordinate invariance when one chooses a specific coordinate representation of given phenomenon in spacetime, in the configuration space ${\cal M}_\text{fields}$, explicit conformal form-invariance is lost once we choose a specific gauge.}


\subsection{GR gauge}


There are simple choices that lead to distinguished gauges. For example, the choice $\hat\phi=M^2_\text{pl}$ leads to the GR gauge. Actually, if we substitute this gauge choice in \eqref{aact-gct}, \eqref{aact-tot-lag'}, we get

\begin{align} {\cal L}_\text{tot}=\frac{\sqrt{-\hat g}M^2_\text{pl}}{2}\left(\hat R-2\lambda M^2_\text{pl}\right)+{\cal L}_m(\hat\chi,\hat\der\hat\chi,\hat g_{\mu\nu}),\label{gr-lag}\end{align} from where the standard GR-EOM are obtained. Notice that $\hat g_{\mu\nu}$ is the physical metric in the GR gauge, so we call it the ``GR metric''. From \eqref{aact-gct} it follows that the metric $g_{\mu\nu}$ in an arbitrary gauge and the GR metric are related by:

\begin{align} g_{\mu\nu}=\left(\frac{M^2_\text{pl}}{\phi}\right)\,\hat g_{\mu\nu}.\label{gr-arb-met}\end{align} Hence, if we know some GR solution, which we label with the index ``$A$'' ($A=1,2,...,N_s$, where $N_s$ is the total number of possible solutions): $\hat g^{(A)}_{\mu\nu}=\hat g^{(A)}_{\mu\nu}(x)$, we don't need to solve the EOM of JFBD-STG theory to get $g^{(A)}_{\mu\nu}=g^{(A)}_{\mu\nu}(x)$. We only need to choose some function $\phi=\phi(x)$ and substitute it in \eqref{gr-arb-met}:

\begin{align} g^{(A)}_{\mu\nu}(x)=\left(\frac{M^2_\text{pl}}{\phi(x)}\right)\,\hat g^{(A)}_{\mu\nu}(x).\nonumber\end{align} For the mass parameter in an arbitrary gauge, we obtain $m=\kappa\sqrt\phi$, where the dimensionless constant $\kappa=\hat m/M_\text{pl}$, where $\hat m$ is the constant mass in GR theory.

However, there is a problem with this gauge: The inverse transformation of \eqref{aact-gct} applied to \eqref{gr-lag} does not lead to \eqref{aact-tot-lag} but to its particular case when $\sigma=0$; that is,

\begin{align} {\cal L}_\text{tot}=\frac{\sqrt{-g}}{2}\left[\phi R+\frac{3}{2\phi}(\der\phi)^2-2\lambda\phi^2\right]+{\cal L}_m(\chi,\der\chi,g_{\mu\nu}).\label{sing-tot-lag}\end{align} This means that the GR gauge does not belong to the conformal equivalence class of the JFBD-STG theory \eqref{aact-tot-lag} for any coupling function $\sigma=\sigma(\phi)$, but only for $\sigma=0$ $\Rightarrow\omega=-3/2$. Notice that this value of the coupling function is singular, as this is the only constant value that is not transformed by the conformal transformation \eqref{sigma-t}.

Let

\begin{align} \mathfrak{G}^A_\text{gr}\equiv\mathfrak{G}^A_0:\left[\phi_0=M^2_\text{pl}, g^{(0,A)}_{\mu\nu}=g^{(0,A)}_{\mu\nu}(t,\vec{x}), \chi_0=\chi_0(t,\vec{x}), \sigma_0\to\infty, m_0=\kappa\,M_\text{pl}\right],\label{gr-ggs}\end{align} denote the GR global gravitational state labeled ``$A$'', where different values $A=1$, $A=2$,...,$A=N_s$ represent different solutions of the GR-EOM: $G^{(0)}_{\mu\nu}=T^{(0,m)}_{\mu\nu}/M^2_\text{pl}$, and let 

\begin{align} \mathfrak{G}^A_j:\left[\phi_j=\phi_j(t,\vec{x}), g^{(j,A)}_{\mu\nu}=g^{(j,A)}_{\mu\nu}(t,\vec{x}), \chi_j=\chi_j(t,\vec{x}), \sigma_j=0, m_j=\kappa\,\phi_j(t,\vec{x})\right],\label{j-gauge}\end{align} denote another arbitrary gauge, labeled ``$j$''. We can relate the GGS ``$j$' and the GR-GGS by means of the following relationships:

\begin{align} g^{(j,A)}_{\mu\nu}(t,\vec{x})=\left(\frac{M^2_\text{pl}}{\phi_j(t,\vec{x})}\right)\,g^{(0,A)}_{\mu\nu}(t,\vec{x}),\;\chi_j(t,\vec{x})=\left(\frac{M^2_\text{pl}}{\phi_j(t,\vec{x})}\right)^\frac{w_\chi}{2}\chi_0(t,\vec{x}),\;\sigma_j=\left(\frac{\der M^2_\text{pl}}{\der\phi_j}\right)^2\left(\frac{\phi_j}{M^2_\text{pl}}\right)^2\sigma_0=0.\nonumber\end{align} Hence, these GGSs belong to the same conformal equivalence class (CEC):

\begin{align} \mathfrak{C}^A_{\sigma=0}=\{\mathfrak{G}^A_0,\mathfrak{G}^A_1,...,\mathfrak{G}^A_j,...,\mathfrak{G}^A_N\},\;\;N\to\infty,\label{3/2-cec}\end{align} where the general element of the class is given by \eqref{j-gauge}. The metric $g^{(j,A)}_{\mu\nu}(t,\vec{x})$ is a solution of the EOM derived from \eqref{sing-tot-lag}, once we choose a function $\phi_j(t,\vec{x})$. Recall that for $\sigma=0$, the KG-type equation \eqref{kg-eom'}: $2\sigma\nabla^2\phi+\sigma_{,\phi}(\der\phi)^2=0$, becomes an identity $0=0$, so the scalar field does not obey an independent EOM; that is, $\phi$ can be chosen at will. In fact, we do not need to solve the EOM resulting from \eqref{sing-tot-lag}. Given the relationship between the GR solution $g^{(0,A)}_{\mu\nu}(t,\vec{x})$ and the metric in another arbitrary gauge in $\mathfrak{C}^A_{\sigma=0}$:

\begin{align} g^{(j,A)}_{\mu\nu}(t,\vec{x})=\left(\frac{M^2_\text{pl}}{\phi_j(t,\vec{x})}\right)\,g^{(0,A)}_{\mu\nu}(t,\vec{x}),\label{rel-met}\end{align} once we choose the function $\phi=\phi_j(t,\vec{x})$, we can determine the metric $g^{(j,A)}_{\mu\nu}$ from a known GR solution $g^{(0,A)}_{\mu\nu}(t,\vec{x})$. Both solutions are physical since they are related to two different (global) gravitational states $\mathfrak{G}^A_0$ and $\mathfrak{G}^A_j$, respectively. Both can describe gravitational phenomena. Only an experiment can make a decisive choice and pick one specific GGS as the one that provides the closest description of a given gravitational phenomenon, among the infinite number of them in $\mathfrak{C}^A_{\sigma=0}$.


One of the most important results here is that the GR gauge belongs to the conformal equivalence class \eqref{3/2-cec} of JFBD-STG theory with a constant coupling parameter $\sigma=0$ ($\omega=-3/2$). That is, although the GR theory itself is not manifestly conformal form-invariant (it should not be because GR is a particular gauge), it belongs to the CEC of conformal form-invariant JFBD-STG theory with the vanishing coupling parameter $\sigma=0$.


\subsection{BD gauge}


The Brans-Dicke theory is a particular case when in \eqref{aact-tot-lag'} we set $\hat\sigma$ to a constant $\hat\sigma=\hat\sigma_0=\omega_{BD}+3/2\neq 0$, where $\omega_{BD}$ is the BD coupling constant. To understand how the gauge choice is forced in this case, let us write the KG-type equation \eqref{kg-eom'}:

\begin{align} 2\hat\sigma\hat\nabla^2\hat\phi+\hat\sigma_{,\hat\phi}(\hat\der\hat\phi)^2=0\;\Rightarrow\hat\nabla^2\hat\phi=0.\nonumber\end{align} As seen, the choice $\hat\sigma=\hat\sigma_0=$ const. leads to the wave equation $\hat\nabla^2\hat\phi=0$ in curved space with metric tensor $\hat g_{\mu\nu}$. The latter equation must be taken together with the independent Einstein-type equations \eqref{einst-eom};

\begin{align} \hat G_{\mu\nu}-\frac{\hat\sigma-3/2}{\hat\phi^2}\left[\der_\mu\hat\phi\der_\nu\hat\phi-\frac{1}{2}\hat g_{\mu\nu}(\hat\der\hat\phi)^2\right]-\frac{1}{\phi}\left(\nabla_\mu\nabla_\nu-g_{\mu\nu}\nabla^2\right)\phi+\lambda\hat\phi\hat g_{\mu\nu}=\frac{1}{\hat\phi}\,\hat T^{(m)}_{\mu\nu}.\nonumber\end{align} Suppose that $\hat\phi_{A'}=\hat\phi_{A'}(t,\vec{x})$ and $\hat g^{(A')}_{\mu\nu}=\hat g^{(A')}_{\mu\nu}(t,\vec{x})$ are a solution of the above system of EOM. We can say that the choice $\hat\sigma=\hat\sigma_0$ forced the scalar field to take the form $\hat\phi_{A'}(t,\vec{x})$. This is what we call here gauge choice, which is a bit different from the singular case where $\sigma=\hat\sigma=0$. For any other gauge choice $\hat\sigma=\hat\sigma(\hat\phi)$, the situation is similar.


The conformal transformation \eqref{aact-gct} from an arbitrary gauge to the BD gauge includes the transformation of the parameter $\sigma$ that can be written as:

\begin{align} \sqrt{\omega_{BD}+3/2}\,\frac{d\hat\phi}{\hat\phi}=\frac{\sqrt{\sigma(\phi)}}{\phi}\,d\phi,\nonumber\end{align} whose straightforward integration yields to:

\begin{align}\hat\phi=C\exp{\left(\frac{1}{\sqrt{\omega_{BD}+3/2}}\int\frac{\sqrt{\sigma(\phi)}}{\phi}\,d\phi\right)},\label{bd-gauge}\end{align} where $C$ is an integration constant. Notice that a choice $\sigma=\sigma(\phi)$ leads to a unique function $\hat\phi=\hat\phi(\phi)$. Notice also that in the limit $\omega_{BD}\to\infty$, from \eqref{bd-gauge} we get $\hat\phi\to$ const., which is GR theory. However, as already explained, the inverse of the generalized conformal transformation \eqref{aact-gct} does not lead to BD theory with an arbitrary $\omega_{BD}$, but to the singular case where $\omega_{BD}=-3/2$ ($\hat\sigma=0$). Hence, the GR gauge does not belong in the conformal equivalence class of the BD theory with an arbitrary coupling parameter. This is given by the following total Lagrangian density:

\begin{align} \mathfrak{C}^{A'}_{\sigma\neq 0}=\{\mathfrak{G}^{A'}_{0'},\mathfrak{G}^{A'}_{1'},...,\mathfrak{G}^{A'}_{j'},...,\mathfrak{G}^{A'}_{N'}\},\;\;N'\to\infty,\label{sigma-cec}\end{align} where

\begin{align} \mathfrak{G}^{A'}_{0'}:\left[\phi^{A'}_{0'}=\phi^{A'}_{0'}(t,\vec{x}), g^{(0',A')}_{\mu\nu}=g^{(0',A')}_{\mu\nu}(t,\vec{x}), \chi_{0'}=\chi_{0'}(t,\vec{x}), \sigma_{0'}=\omega_{BD}+3/2, m^{A'}_{0'}=\kappa\,\phi^{A'}_{0'}(t,\vec{x})\right],\label{0'-gauge}\end{align} represents some solution ``$A'$'' of BD theory; that is, a global gravitational state of BD theory denoted $A'$. The general element of the class is given by:

\begin{align} \mathfrak{G}^{A'}_j:\left[\phi_j=\phi_j(t,\vec{x}), g^{(j,A')}_{\mu\nu}=g^{(j,A')}_{\mu\nu}(t,\vec{x}), \chi_j=\chi_j(t,\vec{x}), \sigma_j=\sigma(\phi_j), m_j=\kappa\,\phi_j(t,\vec{x})\right],\label{j'-gauge}\end{align} Notice that the conformal classes $\mathfrak{C}^A_{\sigma=0}$ defined in \eqref{3/2-cec} and $\mathfrak{C}^{A'}_{\sigma\neq 0}$ given by \eqref{sigma-cec}, are disjoint classes $\mathfrak{C}^A_{\sigma=0}\cap\mathfrak{C}^{A'}_{\sigma\neq 0}=\emptyset$.


\subsection{``Many-worlds'' interpretation of gauge freedom}


Every global gravitational state or gauge requires complete knowledge of the metric and of all fields coupled to gravity at every spacetime point and at all times, so we may conceive of a GGS as the entire history of the universe. Within the conformal equivalence classes $\mathfrak{C}^A_{\sigma=0}$ and
$\mathfrak{C}^{A'}_{\sigma\neq 0}$, there are as many possible histories as there are global gravitational states. Since every GGS is physically meaningful and provides a potential description of specific gravitational phenomena, we refer to the resulting framework as the many-worlds interpretation of gauge freedom \cite{quiros-prd-2025, quiros-arxiv2-2025, quiros-prd-2023}. In a classical context, only one of the GGSs correctly describes the given gravitational phenomenon: the one that more closely reproduces the observational data. However, in the quantum gravitational framework, the $N$ gauges in $\mathfrak{C}^A_{\sigma=0}$ contribute to the probability amplitude $\vphi(A)$ of the solution ``$A$'', while the $N'$ gauges in $\mathfrak{C}^{A'}_{\sigma\neq 0}$ contribute to the probability amplitude $\vphi(A')$ of the solution ``$A'$''. The probability amplitude of some solution can be expressed as \cite{hawking-2, hawking-3, hawking-4}: 

\begin{align} \vphi=\int {\cal D}\phi\,{\cal D}g\,\exp{iS_\text{stg}[\phi,g]},\label{path-int}\end{align} where ${\cal D}\phi$ and ${\cal D}g$, are some measures in the space of all scalar fields $\phi=\phi(x)$ and metrics $g_{\mu\nu}=g_{\mu\nu}(x)$, respectively, and $S_\text{stg}[\phi,g]$ is the classical action of JFBD-STG theory (for simplicity let us avoid any matter contribution):

\begin{align} S_\text{stg}[\phi,g]=\frac{1}{2}\int_{{\cal V}_4}d^4x\sqrt{-g}\left[\phi R-\frac{\omega(\phi)}{\phi}(\der\phi)^2-2\lambda\phi^2\right]+\int_{\der{\cal V}_4}d^3x\sqrt{-h}\,\phi K,\label{jfbd-stg-action}\end{align} where ${\cal V}_4$ is some integration volume, $h_{\mu\nu}=g_{\mu\nu}\pm n_\mu n_\nu$ is the metric induced on the boundary of the integration volume  $\der{\cal V}_4$, which is orthogonal to the unit vector field $n_\mu$, and $K=h^{\mu\nu}K_{\mu\nu}$ ($K_{\mu\nu}=h^\lambda_{\;\mu}h^\sigma_{\;\nu}\nabla_\lambda n_\sigma$ is the extrinsic curvature tensor of the boundary). However, for our qualitative discussion, it will be more appropriate to work with the following sum

\begin{align} \vphi(A)=\lim_{N\to\infty}\sum_{k=1}^N\vphi(A|k),\;\;\vphi(A|k)=\frac{\vphi_0}{N}\exp iS_\text{stg}[\phi_k,g_{(k,A)}],\label{p-amp}\end{align} where $\vphi(A|k)$ is the probability amplitude of the gauge ``$k$'' corresponding to the GGS ``$A$'', $\vphi_0/N$ is some normalization constant, and the sum is over all gauges in the conformal class $\mathfrak{C}^A_{\sigma=0}$. Similar expressions hold for the probability amplitude of solution ``$A'$'', where the sum is over all GGSs in $\mathfrak{C}^A_{\sigma\neq 0}$. Since the overall action \eqref{jfbd-stg-action} is form-invariant under generalized conformal transformations \eqref{gen-gauge-t'}, each probability amplitude contributes not only the same amplitude but also the same phase to the overall probability amplitude. This means that we can choose any single element of the conformal equivalence class, say $k=r$, so the following equality takes place: $\vphi(A)=\vphi_0\exp iS_\text{stg}[\phi_r,g_{(r,A)}]$. However, we know that every possible gravitational state, regardless of whether it is a solution of the EOM, contributes to the path integral. Yet, the classical action \eqref{jfbd-stg-action} itself is conformal form-invariant, so the above result is independent of the derived EOM (perhaps the normalization constant must be modified).

In the literature, one finds contradictory opinions about the role of conformal symmetry in quantum-gravitational interactions. From the path integral viewpoint on quantum gravity, since the classical action \eqref{jfbd-stg-action} is conformal form-invariant, we found that this symmetry may play a role. It is only required to define a suitable conformal invariant measure ${\cal D}\phi\,{\cal D}g$ in the configuration space.



\section{Consequences of conformal symmetry for scalar-tensor gravitational theories}
\label{sect-extra}


In this section, we summarize the outstanding results and comment on the main phenomenological consequences of our current framework, in which STG theories in the JFBD parametrization are treated as conformal form-invariant, and the active approach to CTs is adopted. The better way to put our results in perspective is to compare them with the well-known results in the bibliography on STG theory. Below, we list our main results, accompanied by a brief comparison with established knowledge, and then discuss the phenomenological consequences of these results.

\begin{enumerate}

\item[(a)]{\bf The minimal coupling of matter field to the metric is not transformed by the generalized conformal transformations.} The assumption that, under CTs: $\hat g_{\mu\nu}=\Omega^2g_{\mu\nu}$, $\hat\phi=\Omega^{-1}\phi$, $\hat\chi=\Omega^{w_\chi}\chi$, the mass parameters transform as $\hat m=\Omega^{-1}m$, inevitably leads to the Lagrangian density of matter fields, including that of perfect fluids, being form-invariant under the conformal transformations \cite{quiros-prd-2025}; ${\cal L}_m(\chi,\der\chi,g_{\mu\nu})={\cal L}_m(\hat\chi,\hat\der\chi,\hat g_{\mu\nu})$. This is in contrast with the well-known result \cite{deruelle_2011, chiba_2013, sasaki-2015}, that $\sqrt{-g}\,L_m(\chi,\der\chi,g_{\mu\nu})=\sqrt{-\hat g}\,\Omega^{-4}L_m(\chi,\der\chi,\Omega^{-2}\hat g_{\mu\nu})$; that is, in the conformal frame the matter fields are minimally coupled to the metric $g_{\mu\nu}$ and not to $\hat g_{\mu\nu}$. Moreover, this known result implies that matter fields are not transformed by conformal transformations ($\chi\to\chi$), which is not generally true: only null fields and radiation are not transformed by the CTs, as opposed to fermions, for example, which are indeed transformed. We have shown that, if we assume that $m\to\Omega^{-1}m$ under CTs, this result is incorrect since $\sqrt{-g}\,L_m(\chi,\der\chi,g_{\mu\nu})=\sqrt{-\hat g}\,L_m(\hat\chi,\hat\der\hat\chi,\hat g_{\mu\nu})$. Therefore, the conformal matter fields $\hat\chi$ are minimally coupled to the conformal metric $\hat g_{\mu\nu}$. In other words, the minimal coupling of matter fields to the metric is not modified by the conformal transformations. Despite the apparent triviality of the result, to our knowledge, it has not been discussed before in the literature on STG theory in the JFBD parametrization. 

\item[(b)]{\bf STG theory in the JFBD parametrization is conformal form-invariant under generalized conformal transformations; that is, Dicke's principle is correct.} The overall Lagrangian density of JFBD-STG theory;

\begin{align} {\cal L}_\text{tot}=\frac{\sqrt{-g}}{2}\left[\phi R-\frac{\omega(\phi)}{\phi}(\der\phi)^2-2\lambda\phi^2\right]+{\cal L}_m(\chi,\der\chi,g_{\mu\nu}),\label{jfbd-stg-tot-lag}\end{align} is form-invariant under the following generalized conformal transformations: 

\begin{align} g_{\mu\nu}\rightarrow\hat g_{\mu\nu}=\Omega^2g_{\mu\nu},\;\phi\rightarrow\hat\phi=\Omega^{-2}\phi,\;\chi\rightarrow\hat\chi=\Omega^{w_\chi}\chi,\nonumber\\
\omega\rightarrow\hat\omega=\frac{\omega+6\frac{\Omega_{,\phi}}{\Omega}\phi\left(1-\frac{\Omega_{,\phi}}{\Omega}\phi\right)}{\left(1-2\frac{\Omega_{,\phi}}{\Omega}\phi\right)^2},\;m\rightarrow\hat m=\Omega^{-1}m,
\label{gct-t}\end{align} that include a suitable transformation of the coupling function $\omega=\omega(\phi)$ and of the mass parameter $m=m(\phi)$. Form-invariance of the gravitational part of ${\cal L}_\text{tot}$ under GCTs, that is, the vacuum case, was discussed for the first time in \cite{faraoni_1998}. This result, with the inclusion of the result in item (a), generalizes the latter one by including matter fields coupled to gravity.

\item[(c)]{\bf The Ward identity for the matter Lagrangian density cannot be ignored since it is a consequence of conformal form-invariance of the matter Lagrangian density.} It is not enough to show form-invariance of the overall Lagrangian density \eqref{jfbd-stg-tot-lag} under GCT \eqref{gct-t}, to have a consistent conformal invariant gravitational theory. Additionally, it is required to take into account the Ward identity

\begin{align} g^{\mu\nu}\frac{\delta{\cal L}_m}{\delta g^{\mu\nu}}=-\phi\frac{\delta{\cal L}_m}{\delta\phi}\;\Rightarrow\;\frac{\delta{\cal L}_m}{\delta\phi}=\frac{\sqrt{-g}}{2\phi}\,T^{(m)},\label{ward-id-m}\end{align} due to conformal form-invariance of the matter Lagrangian density. Fulfillment of the Ward identity \eqref{ward-id-m} is inevitable if one assumes that the mass parameter transforms as $m\to\Omega^{-1}m$, under CTs. In this case, one can write $m\propto\sqrt\phi$, so the Lagrangian density of timelike matter fields depends on $\phi$; that is, $\delta{\cal L}_m/\delta\phi\neq 0$. This has been a missing element in the well-known studies of SGT theories in the literature \cite{fujii_book, faraoni_book, quiros_ijmpd_2019}. The consequence has been an incorrect derivation of the KG-type EOM of the scalar field \eqref{kgbd-eom'}; $\left(3+2\omega\right)\nabla^2\phi+\omega_{,\phi}(\der\phi)^2=2\left(\phi V_{,\phi}-2V\right)+T^{(m)}$. The correct EOM \eqref{kg-eom} reads as,\footnote{For radiation and massless matter fields, under the assumption of a conformal symmetric potential $V=\lambda\,\phi^2$, equation \eqref{kg-eom'} is obtained. The novel result is that, if we account for the Ward identity for matter, the latter EOM is obtained regardless of the presence of null or timelike fields.} $2\omega\nabla^2\phi+\left(\omega_{,\phi}-\omega/\phi\right)(\der\phi)^2+\phi R=2\phi V_{,\phi}-T^{(m)}$, or if take into account the trace of Einstein-type EOM \eqref{einst-eom}, which is given by \eqref{trace-eom}: $3\nabla^2\phi+\omega(\der\phi)^2/\phi-\phi R=T^{(m)}-4V$, we obtain the equivalent KG-type EOM \eqref{kg-eom'}: $(2\omega+3)\nabla^2\phi+\omega_{,\phi}(\der\phi)^2=0$. It does not depend on the curvature of spacetime nor on the (trace of the) matter SET. For BD theory, where $\omega=\omega_{BD}=$ const., the above EOM is just the wave equation for the scalar field propagating in a curved background; $\nabla^2\phi=0$. The dynamics of the scalar field, being independent of the curvature of space and of its matter content, is a direct consequence of form-invariance under GCTs \eqref{gct-t}, which is tightly linked to gauge freedom in the present framework.

\item[(d)]{\bf Conformal form-invariance of STG theory in JFBD parametrization inevitably leads to a ``dark'' fifth force.} This item is closely related to the former one, since what we shall discuss is also a direct consequence of the Ward identity \eqref{ward-id-m} for the matter Lagrangian density. If we find the covariant divergence of the Einstein-type EOM \eqref{einst-eom}, we find that;

\begin{align} \nabla^\lambda T^{(m)}_{\lambda\mu}=-\frac{\der_\mu\phi}{2\phi}\left[2\omega\nabla^2\phi+\left(\omega_{,\phi}-\frac{\omega}{\phi}\right)(\der\phi)^2+\phi R-2\phi V_{,\phi}\right].\label{div-set}\end{align} If we insert in the RHS of the above equation the well-known (mostly incorrect) KGBD-type EOM, which is exclusively the one we found in the literature: $2\omega\nabla^2\phi+\left(\omega_{,\phi}-\omega/\phi\right)(\der\phi)^2+\phi R=2\phi V_{,\phi}$, we obtain the standard conservation equation $\nabla^\lambda T^{(m)}_{\lambda\mu}=0$. This equation is not conformal invariant, of course. In contrast, if we take into account the Ward identity \eqref{ward-id-m}, the KGBD-type EOM obtained is given by \eqref{kg-eom}, which, when substituted into the RHS of \eqref{div-set}, leads to the nonhomogeneous continuity equation \eqref{nhom-cont-eq}:

\begin{align} \nabla^\lambda T^{(m)}_{\lambda\mu}=\frac{\der_\mu\phi}{2\phi}\,T^{(m)},\nonumber\end{align} which is a conformal form-invariant equation. Notice that, for radiation, since $T^{(m)}=0$, the standard conservation equation is obtained. For point-like timelike and null fields, the equivalent relationships are given by EOM \eqref{non-geod} and \eqref{0-geod};

\begin{align} \frac{d^2x^\alpha}{ds^2}+\left\{^\alpha_{\mu\nu}\right\}\frac{dx^\mu}{ds}\frac{dx^\nu}{ds}=\frac{\der_\mu \phi}{2\phi}h^{\mu\alpha},\;\frac{dk^\mu}{d\xi}+\left\{^\mu_{\nu\sigma}\right\}k^\nu k^\sigma=0,\nonumber\end{align} respectively. These are clearly conformal form-invariant equations. The most important phenomenological consequence of this result is the occurrence of a fifth force $f_{(5)}^\alpha=h^{\alpha\mu}\der_\mu\phi/2\phi$, which is required to preserve conformal form-invariance. This will be discussed below. 

\item[(e)]{\bf Only the active approach to conformal transformations is suitable to investigate the phenomenological consequences of conformal symmetry.} Only the active approach to CT has physical and phenomenological consequences. Ironically, the passive approach, which has been used by far most frequently in the literature, is not a suitable framework for discussing conformal symmetry. According to PACT, conformal transformations relate different representations: ${\cal R}_g:(g_{\mu\nu},\phi,\chi)$ and $\hat{\cal R}_g:(\hat g_{\mu\nu},\hat\phi,\hat\chi)$, of the same GGS in the configuration space ${\cal S}_\mathfrak{g}:(\mathfrak{g}_{\mu\nu},\Psi)$ in ${\cal M}_\text{fields}$, where the physically meaningful fields; that is, those that are invariant under CTs, can be defined as in \eqref{inv-met};

\begin{align} \mathfrak{g}_{\mu\nu}=\frac{\phi}{M^2_\text{pl}}\,g_{\mu\nu},\;\Psi=\left(\frac{\phi}{M^2_\text{pl}}\right)^\frac{w_\chi}{2}\chi.\nonumber\end{align} Thanks to the equality \eqref{mat-lag-equal}; ${\cal L}_m(\chi,\der\chi,g_{\mu\nu})={\cal L}_m(\Psi,\mathfrak{d}\Psi,\mathfrak{g}_{\mu\nu})$, and, after rewriting the gravitational Lagrangian density in terms of the conformal invariant fields, the total Lagrangian density \eqref{jfbd-stg-tot-lag} takes the following covariant form:

\begin{align} {\cal L}_\text{tot}=\frac{\sqrt{-\mathfrak{g}}}{2}\left\{\mathfrak{R}-(\mathfrak{D}\Phi)^2-2\lambda\right\}+{\cal L}_m(\Psi,\mathfrak{d}\Psi,\mathfrak{g}_{\mu\nu}),\label{pact-stg-tot-lag}\end{align} which is the GR Lagrangian density with background matter fields $\Phi\equiv\int\sqrt{\sigma(\phi)}d\phi/\phi$ and $\Psi$. All fields appearing in \eqref{pact-stg-tot-lag} are physical; that is, conformal invariants, so a conformal transformation in this equation amounts to the identity transformation $\mathfrak{g}_{\mu\nu}\to\mathfrak{g}_{\mu\nu}$, $\mathfrak{R}\to\mathfrak{R}$, $\Phi\to\Phi$, and $\Psi\to\Psi$. The auxiliary fields $g_{\mu\nu}$, $\phi$, $\chi$, $\omega$, and $m$,\footnote{According to PACT, the mass parameter in the matter Lagrangian density, must be replaced by the conformal invariant mass $\mathfrak{m}=\left(M_\text{pl}/\sqrt\phi\right)\,m$.} which suffer the generalized conformal transformations \eqref{gct-t}, have been removed from the covariant description \eqref{pact-stg-tot-lag}. Hence, since GR itself is not manifestly conformal form-invariant, this approach is not suitable for investigating the conformal symmetry.

In contrast, according to AACT, the fields $g_{\mu\nu}$, $\phi$, $\chi$, $\omega$, and $m$, which suffer the generalized conformal transformations \eqref{gct-t}, are physical since they define a global gravitational state, while $\hat g_{\mu\nu}$, $\hat\phi$, $\hat\chi$, $\hat\omega$, and $\hat m$, define another, different, GGS. In this case, GCT \eqref{gct-t} relates two different gravitational states. Therefore, the AACT is suitable for investigating the physical and phenomenological consequences of conformal symmetry. The discussion of the complementary approaches to CT has no parallel in the established literature. This subject has been formerly discussed only in \cite{quiros-prd-2025, quiros-arxiv2-2025}.

\item[(f)]{\bf Gauge freedom and the active approach to the conformal transformations lead to the ``many-worlds'' representation of conformal invariance.} If we adopt the active approach to CT, gauge freedom in JFBD-STG theory leads to the gravitational many-worlds picture. This means that every GGS and its conformal gravitational states provide suitable representations of given gravitational phenomena. Classical gravitational experiments pick out a single GGS among the infinity of them in a conformal equivalence class: the gauge that more closely reproduces the experimental data. In contrast, in quantum gravitational phenomena, every GGS counts. In this case, due to conformal form-invariance of the total action, every GGS contributes the same amplitude and phase to the overall probability amplitude of those phenomena. The gravitational many-worlds picture has been discussed in the literature only in \cite{quiros-prd-2025, quiros-arxiv2-2025, quiros-prd-2023}. In the context of JFBD-STG theories, it is discussed here for the first time.

\item[(g)]{\bf General relativity and the Brans-Dicke theory do not belong to the same conformal equivalence class; that is, there is no relationship between the two.} An important result shows that the GR gauge belongs to the conformal equivalence class $\mathfrak{C}^A_{\sigma=0}$ \eqref{3/2-cec} of JFBD-STG theory with a constant coupling parameter $\sigma=0$ ($\omega=-3/2$). That is, although the GR theory itself is not manifestly conformal form-invariant (it should not be because GR is a particular gauge), it belongs to the CEC of conformal form-invariant JFBD-STG theory with the vanishing coupling parameter $\sigma=0$. Similarly, BD theory is a gauge in the CEC $\mathfrak{C}^A_{\sigma\neq 0}$ \eqref{sigma-cec}. Since these are disjoint classes, GR cannot be obtained from BD theory, and vice versa.

\end{enumerate}

A phenomenological consequence of the above results, in particular of result (c), is that the screening chameleon mechanism \cite{cham-1, cham-2} does not occur in conformal invariant STG theories. Actually, the chameleon field must obey the following EOM \cite{cham-quiros}:

\begin{align} \nabla^2\phi=\frac{\der V_\text{eff}}{\der\phi},\;V_\text{eff}=\frac{2}{2\omega_{BD}+3}\left[\phi V(\phi)-3\int d\phi V(\phi)+\frac{\phi}{2}\,T^{(m)}\right],\label{cham-eom}\end{align} where the effective potential depends on the surrounding background matter through the trace $T^{(m)}$ of the matter SET. This dependence is central to the chameleon mechanism in STG theories. In contrast, the KG-type EOM \eqref{kg-eom'} is unrelated to the matter content, the background curvature, and the conformal invariant self-interaction potential $V(\phi)=\lambda\,\phi^2$.


Another phenomenological consequence of this result is that the gravitational constant measured in Cavendish experiments is not affected by the coupling parameter $\omega(\phi)$, as we shall show below. In contrast, according to the usual understanding of STG theories in the literature, the measured Newton constant is given by \cite{brans-dicke-1961, fujii_book, quiros_ijmpd_2019}:

\begin{align} 8\pi G_N=\frac{1}{\phi_0}\left(\frac{3+2\omega_0+e^{-M_0 r}}{3+2\omega_0}\right),\label{bd-newton-c}\end{align} where $\phi_0$ is the background value of the BD scalar field, $\omega_0=\omega(\phi_0)$, and $M_0=\sqrt{\phi_0 V_{,\phi\phi}(\phi_0)/(3+2\omega_0)}$ is the mass of the propagating scalar perturbation. The formal limits $M_0\to\infty$ and $M_0\to 0$ lead to general relativity and to BD theory, respectively.

For simplicity of the analysis, to show that $G_N$ is not affected by $\omega(\phi)$ in the present framework, where STG theory in the JFBD parametrization is conformal form-invariant, let us assume a vanishing self-interaction potential. We work in the weak-field limit; that is, we consider the linearization of the fields and of the EOM around the small perturbations $h_{\mu\nu}$ and $\sigma$;

\begin{align} g^L_{\mu\nu}(x)=\eta_{\mu\nu}+h_{\mu\nu}(x),\;\phi_L(x)=\phi_0+\sigma(x),\label{perts}\end{align} where $\phi_0=(8\pi G_N)^{-1}$ and $\eta_{\mu\nu}=\text{diag}(-1,1,1,1)$ is the Minkowski metric. We require that at infinity both $h_{\mu\nu}\to 0$ and $\sigma\to 0$. We also consider the Newtonian limit (matter objects are at rest), so $T^{(m)}_{00}=\rho(x)$, $T^{(m)}_{ij}=0$, and $T^{(m)}=-\rho(x)$. The linearized LHS of EOM \eqref{einst-eom} reads as

\begin{align} \phi_L{\cal E}^L_{\mu\nu}=\frac{\phi_0}{2}\left[-\nabla^2h_{\mu\nu}+\der_\mu\der_\lambda h^\lambda_{\;\;\nu}+\der_\nu\der_\lambda h^\lambda_{\;\;\mu}-\eta_{\mu\nu}\der_\lambda\der_\rho h^{\lambda\rho}-\left(\der_\mu\der_\nu-\eta_{\mu\nu}\nabla^2\right)h-\frac{2}{\phi_0}\left(\der_\mu\der_\nu-\eta_{\mu\nu}\nabla^2\right)\sigma\right],\label{lin-1}\end{align} where $\nabla^2=\eta^{\mu\nu}\der_\mu\der_\nu$, and $h=\eta^{\mu\nu}h_{\mu\nu}=h^\lambda_{\;\;\lambda}$. If we introduce the new field, 

\begin{align} \psi_{\mu\nu}=h_{\mu\nu}-\frac{1}{2}\eta_{\mu\nu}\,h-\frac{1}{\phi_0}\,\eta_{\mu\nu}\,\sigma,\label{lin-2}\end{align} so the mixing between the metric and scalar field perturbations in \eqref{lin-1} is removed:

\begin{align} \phi_L{\cal E}^L_{\mu\nu}=\frac{\phi_0}{2}\left(-\nabla^2\psi_{\mu\nu}+\der_\mu\der_\lambda\Psi^\lambda_{\;\;\nu}+\der_\nu\der_\lambda\Psi^\lambda_{\;\;\mu}-\eta_{\mu\nu}\der_\rho\der_\lambda\psi^{\lambda\rho}\right).\nonumber\end{align} Next, thanks to the coordinate choice freedom, we introduce the following coordinate conditions: $\der_\lambda\psi^\lambda_{\;\;\mu}=0$. The linearized Einstein-type EOM \eqref{einst-eom} then reads as

\begin{align} \nabla^2\psi_{\mu\nu}=-\frac{2}{\phi_0}\,T^{(m)}_{\mu\nu}\;\Rightarrow\;\nabla^2\psi=-\frac{2}{\phi_0}\,T^{(m)}.\label{lin-3}\end{align} 

The key difference between conventional JFBD-STG theory and the present conformal invariant version lies in the new KG-type EOM \eqref{kg-eom'}. The linearized version of this EOM reads as: $\nabla^2\sigma=0$. The unique solution of this equation, which is well-behaved at infinity, is $\sigma=0$. This leads to great simplification due to the removal of the scalar field perturbation from the final result. Actually, if we substitute this result in \eqref{lin-2}, and take into account \eqref{lin-3}, we get that

\begin{align} \nabla^2h_{\mu\nu}=-\frac{2}{\phi_0}\left[T^{(m)}_{\mu\nu}-\frac{1}{2}\eta_{\mu\nu}T^{(m)}\right]\;\Rightarrow\;\nabla^2h_{00}=-\frac{1}{\phi_0}\,\rho=-8\pi G_N\,\rho\;\Rightarrow\;h_{00}=2G_N\int d^3x\frac{\rho}{r}.\label{poisson-eq}\end{align} For a point-like source with mass $M$ we have that $h_{00}=2G_NM/r$, where we recognize the standard Newtonian gravitational potential $U=G_NM/r$. 

Notice that the background value $\phi_0$ depends on the spacetime point. In general, the measured Newton constant $G_\text{eff}(x)=1/8\pi\phi(x)=1/8\pi M^2_\text{pl}(x)$, depends on the spacetime point. Yet, as long as the experimental measurements are involved, since the mass unit $u_m(x)\propto\sqrt{\phi(x)}$ depends on the scalar field in the same universal way as any mass parameter $m=\kappa\,\sqrt\phi$, then, the measured value of the Newton constant is the same at any point in spacetime. To see this, it is better to introduce the dimensionless function $\vphi(x):=\phi(x)/M^2_\text{pl}$, where $M_\text{pl}=1.22\times 10^{19}$ GeV. Then, $M_\text{pl}(x)=M_\text{pl}\sqrt{\vphi(x)}$. The mass unit reads as: $u_m(x)=u_0\sqrt{\vphi(x)}$, where $u_0$ is a constant number (for example, $u_0=1$ mass unit). The quantity measured in experiments is the ratio:

\begin{align} \frac{M_\text{pl}(x)}{u_m(x)}=\frac{M_\text{pl}}{u_0}=\text{const.},\nonumber\end{align} hence, the point-dependence of the measured Newton constant $G_\text{eff}(x)\propto M^{-2}_\text{pl}(x)$, cannot be revealed in local (classical) experiments. Only redshift experiments are suitable to check that $G_\text{eff}$ is a function of the spacetime point. This result can be stated in a bit different way: variations of the measured Newton constant in spacetime are compensated by similar variations of the measuring ``stick''. In other words, local experiments cannot put constraints on the free parameters of STG theory in JFBD parametrization, such as the coupling function $\omega(x)$. 

Another interesting phenomenological consequence of conformal form-invariant JFBD-STG theory is related to the inevitability of a fifth force $f_5^\alpha=h^{\alpha\mu}\der_\mu\phi/2\phi$ (see item (d) above). This force is orthogonal to the four-velocity vector $u^\lambda f^{(5)}_\lambda=0$ and acts only on timelike fields. Null fields and radiation do not interact with $f_{(5)}^\alpha$. This suggests that the fifth force may explain (at least in principle) the dark cosmological sector, composed of dark matter and dark energy.


The above-explained consequences of conformal symmetry in JFBD-STG theory, which we call the ``unknown face'' of STG theory, are opposed to the related consequences when conformal symmetry is ignored. In the latter case, we face the well-known STG theory, which includes the BD theory as a particular case. In this case, experiments set tight constraints on the BD coupling constant $\omega_{BD}>10^6$ \cite{w-bound}, the chameleon effect screens the BD field in local experiments, and we face the conformal frame issue. 

Are both descriptions of STG theory in JFBD parametrization, the conformal form-invariant and the standard ones, physically equivalent? A quick answer is: no, they are not physically equivalent. They are not even mathematically equivalent because of different EOMs for the scalar field. But there are other related questions that must be answered. For example, can we ignore a manifest symmetry of the overall action and of the resulting EOM just because we want to? If we can ignore at will a symmetry of the theory and see what happens, then we can ignore the conformal form-invariance of the JFBD-STG theory, so we are faced with the well-known understanding of STG theory in the literature. Unfortunately, any symmetry of the mathematical structure of the theory is inherent in that structure, so we can do nothing about it; that is, we cannot ignore a manifest symmetry of the action and of the derived EOM. Another related question is the following: can we choose how the masses behave under conformal transformations? The answer is yes; however, the choice has consequences that must be properly considered. The transformation of the mass $m$ and of the Planck constant $h$, under CTs, are not independent of each other. The Planck constant has dimensions, $[h]=[M][L]^2[T]^{-1}$, where $[M]$, $[L]$, and $[T]$ are the mass, length, and time dimensions, respectively. Under CTs, the latter dimensions transform as the proper length and the proper time elements do. Consequently, $[L]^2[T]^{-1}\rightarrow\Omega\,[L]^2[T]^{-1}$. Hence, if we assume that the CTs do not transform the Planck constant, the mass parameter $m$ must be transformed as $m\to\Omega^{-1}m$. Otherwise, if we assume that the mass parameter is a constant, then the Planck constant transforms as $h\rightarrow\Omega\,h$. Independent of the choice, the Compton wavelength $\lambda_c=h/m$ transforms as any other length parameter: $\lambda_c\to\Omega\,\lambda_c$. The choice of the transformation of the mass is critical to the description of STG theory in the JFBD parametrization. If we choose that $m\to m$ under CTs, then the Ward identity \eqref{ward-id-m} does not take place and the standard EOM are obtained. In this case, the presence of timelike matter fields breaks the conformal invariance. Except for an issue, this is basically the well-known understanding of STG theory in the literature. The issue is that, in the literature on conformal transformations, the choice of the mass transformation $m\to\Omega^{-1}m$, appearing in Dicke's paper \cite{dicke-1962}, is unanimously adopted. Hence, the matter action is form-invariant under CTs, and the Ward identity cannot be ignored. This leads to a KGBD-EOM which does not coincide with the one appearing in the literature on STG theory. A final question: can we ignore the transformation of the coupling parameter $\omega(\phi)$ under CTs even if $\phi\to\hat\phi=\Omega^{-2}\phi$? This would entail an additional nontrivial bound $\omega(\phi)=\omega(\hat\phi)=\omega(\Omega^{-2}\phi)$. Hence, in general, the transformation of the coupling parameter in \eqref{gct-t} cannot be ignored under any circumstances.



\section{Discussion and Conclusion}
\label{sect-discu}


It is commonly believed that dimensionless quantities and dimensionful parameters, such as mass,\footnote{If we assume that the mass parameter is unchanged by CT \eqref{conf-t}: $m\rightarrow m,$ the inclusion of a mass term in the Lagrangian density, as in \cite{deser_1970}, breaks the conformal symmetry of an otherwise conformal invariant theory.} are not transformed by CT \eqref{conf-t}. This belief is based on the understanding of conformal transformations \eqref{conf-t} as local scale transformations \cite{bekenstein_1980}. However, according to another trend in the bibliography, under CT \eqref{conf-t} the mass parameter transforms as $m\rightarrow\Omega^{-1}m$ \cite{dicke-1962, bekenstein_1980, casas_1992, fujii_book, faraoni_book, quiros_ijmpd_2019}. As discussed in \cite{bekenstein_1980}, local scale transformations are not unit transformations \cite{dicke-1962}; they are active enlargements of a system whose usefulness depends on the absence of a length scale. Meanwhile, GCTs \eqref{gct} are units' transformations whose physical and geometrical meanings follow from the existence of some scale of length. Here, we assume that conformal transformations are unit transformations, so the mass parameter is a point-dependent field transforming as $m\to\Omega^{-1}m$.

The results of the present document on STG theory in JFBD parametrization and conformal symmetry, challenge the standard viewpoint on conformal invariance. In the bibliography, one very often encounters assertions which are similar to the following ones: ``If conformal symmetry is a fundamental symmetry in nature, it must be broken.'' ``Our world is obviously not conformally invariant.'' etc. These assertions are based on several underlying hypotheses. Some of them come from high-energy physics phenomenology, where conformal symmetry (also local scale symmetry) has a very different meaning \cite{jackiw-1972}. Other hypotheses are based on a very rigid understanding of conformal transformations. For example, a hypothesis that is taken as a fait accompli is that the presence of any dimensionful scale, such as the Planck scale $M_\text{pl}$, breaks conformal symmetry. In the present document, we have assumed a different hypothesis: conformal transformations act exclusively on fields. The constants, whether they are dimensionless or dimensionful, or fundamental or not, are not affected by these transformations. Hence, the presence of scales (in the sense that some constants may appear in the Lagrangian density and, consequently, in the derived EOM) in the present framework does not affect conformal symmetry. Similarly, in the present work we assume that the mass parameter, which is usually associated with conformal symmetry breaking, is also a field \cite{hobson-epjc-2022}; $m=m(\phi)=\kappa\sqrt\phi$, which, under CT, transforms as $m\rightarrow\hat m=\Omega^{-1}m$ \cite{dicke-1962, bekenstein_1980, faraoni_prd_2007, casas_1992, fujii_book, faraoni_book}. Hence, nonvanishing masses do not break conformal symmetry \cite{quiros-prd-2025}. This assumption is at the core of the demonstration that the Lagrangian density of fundamental matter fields is conformal form-invariant, leading to the Ward identity \eqref{mat-ward-id}. 

There is another mathematically sound result that has long fueled the belief that conformal invariance must be a broken symmetry in nature. Many years ago in \cite{deser_1970}, working in the context of CGR theory \eqref{aact-lag}, it was demonstrated that, for conformal invariance to be a symmetry of the latter theory, the trace of the stress-energy tensor of matter must vanish. That is, only massless radiation can be coupled to gravity in this theory. However, this result is highly dependent on the assumption made in \cite{deser_1970} that CTs do not transform the (constant) mass parameters. If one assumes point-dependent masses, then vanishing trace of the matter SET in CGR is not required \cite{quiros-arxiv2-2025}. In other words, the result of \cite{deser_1970} rests on the assumption that conformal transformations are just local scale transformations, instead of assuming them as transformations of units \cite{dicke-1962}, as we do here.  

A popular argument against the real physical impact of point-dependent masses and the associated fifth force is that it is the very nature of the conformal transformation that these effects exactly cancel each other out. That is, the trajectories of test masses under a fifth force, as measured with point-dependent clocks and rods, give the same numerical values as in the absence of such modifications. This is true because not only the masses of timelike fields, but also the clocks and rods, which are made of these fields, vary from point to point in the same universal way. However, due to the $\phi$-dependence of the electron mass: $m_e(x)=\kappa_e\sqrt{\phi(x)}$, where $\kappa_e$ is some dimensionless parameter, two identical atoms with atomic number $Z$, located at distant spacetime points, will emit/absorb photons with a slightly different frequency. The lucky fact is that photons follow null geodesics of Riemann geometry (photons and radiation, in general, are blind to the fifth force), so we know how to control the gravitational redshift of frequency, also called curvature redshift. This means that, in principle, we can actually measure the fifth force effect, which originates from the point-dependent property of masses, in redshift experiments such as, for example, the Pound-Rebka-Snider experiments \cite{pound-1960, pound-1964, pound-1965}.


In this document, we introduce a new dimension to the subject of conformal transformations: we apply the notions of passive and active CT to the study of the conformal frame issue. While passive CT is a redundancy of the physical description of a given gravitational state, active CT entails actual changes in the gravitational state itself. In other words, while passive CT has no phenomenological implications, active conformal transformation has physical consequences. This approach to conformal symmetry has been proposed in \cite{quiros-prd-2025}, and here we have applied it within the CFI context. Some researchers might think that the active approach to conformal transformations is merely a mathematical or philosophical artifice without physical consequences. But as we have demonstrated, AACT is a possibility that cannot be simply ignored. It has no parallel in the bibliography, even if it is inspired by the active interpretation of coordinate transformations. Here we treat conformal transformations as ``coordinate transformations'' in the configuration space; that is, unlike standard coordinate transformations in spacetime, the generalized (abstract) coordinates $g_{\mu\nu}$, $\phi$, $\chi$, and $m$, have themselves physical meaning. Recall that, according to the active approach to CTs, any collection $\{g_{\mu\nu}(x),\phi(x),\chi(x),m(x)\}$, being a point in the configuration space, defines a global gravitational state. Any other point $\{\hat g_{\mu\nu}(x),\hat\phi(x),\hat\chi(x),\hat m(x)\}$ in the configuration space, related to the former one through a generalized conformal transformation, defines a different GGS.

One of the goals of this document has been to (try to) convince those readers who think that the active approach to conformal transformations is merely a mathematical or philosophical artifice without physical consequences that the AACT actually carries new physics. In fact, a similar argument can be made for passive conformal transformations. We have shown that the latter transformations act on auxiliary fields (recall that, according to PACT, only conformal invariant quantities matter), which are removed from the covariant formulation, so conformal symmetry can be at most a spurious symmetry of STG theory. Hence, the question is: Does conformal symmetry make any sense at all? If we believe that the answer is positive, then the only way for this symmetry to have some observable effect is that it is associated with an active conformal transformation.


Our investigation of the statement made by Dicke in \cite{dicke-1962} that ``the laws of physics must be invariant under a transformation of units'', concludes that, if we undertake the active approach to CT, Dicke's principle is satisfied by STG theory in JFBD parametrization \eqref{tot-lag}, which is form-invariant under generalized conformal transformations \eqref{gct-t}. If we adopt restricted conformal transformations, that is, the group of transformations \eqref{gauge-t}, which does not contain the transformation of the coupling parameter, form-invariance under \eqref{gauge-t} is not a symmetry of JFBD-STG, so Dicke's principle is not verified. In the latter case, the CFI arises, since the theory does not admit a covariant formulation in the configuration space.


We conclude that the CFI is due to two factors. The first factor is an incorrect omission of the transformation \eqref{w-gauge-t} of the coupling parameter $\omega$, in the conformal transformation group. Since CT includes a transformation of the BD field $\phi\rightarrow\hat\phi=\Omega^{-2}\phi$, any field-dependent parameter, as $\omega=\omega(\phi)$, must also be transformed. Then, instead of restricted transformations \eqref{gauge-t}, we must consider generalized transformations \eqref{gct-t} as the correct group of transformations. The second factor is the ignorance of the conformal form-invariance of the matter action due to the point-dependence of the masses of timelike fields. As shown in \cite{quiros-prd-2025}, for fundamental matter fields as well as for perfect fluids, if the masses transform under CT as $m\rightarrow\Omega^{-1}m$, which means that the energy density of perfect fluids transforms as $\rho\rightarrow\Omega^{-4}\rho$, the Lagrangian density of matter fields and of perfect fluids is form-invariant under CT. This leads to the Ward identity \eqref{mat-ward-id}, which is fundamental for the derivation of the correct conformal form-invariant KG-type EOM.


{\bf Acknowledgments} We thank Sergey Odintsov and Sotirios Karamitsos for pointing out several relevant bibliographic references, and Felipe F Faria for the useful exchange of ideas and critical comments, and also for pointing out several bibliographic references. We also acknowledge FORDECYT-PRONACES-CONACYT for support of the present research under grant CF-MG-2558591.  






\end{document}